\documentclass[12pt]{article}

\pdfoutput=1

\usepackage{graphicx}
\usepackage{amsmath}
\usepackage{color}
\usepackage[table]{xcolor}
\usepackage{multirow}
\begin{document}

%%-move to normal A4-%%
\hoffset = -0.3truecm
\voffset = -1.1truecm

\title{\bf
Multiple bifurcations and transitions for electrically charged monopole-antimonopole chain and vortex-ring solutions}

\author{
{\bf Amin Soltanian,}
{\bf Rosy Teh\footnote{E-mail: rosyteh@usm.my} and Khai-Ming Wong}\\
{\normalsize School of Physics, Universiti Sains Malaysia}\\
{\normalsize 11800 USM Penang, Malaysia}}

\date{December 11, 2014}
\maketitle

\begin{abstract}
The dependence of physical properties of the electrically charged monopole-antimonopole pair (MAP) solutions in the Higgs self-coupling constant is previously investigated. In this paper we study the three-poles monopole-antimonopole chain (MAC) solutions. The study includes $\phi$-winding number $n=2,3,4$, and $5$. For the case of $n=2$, there is no bifurcating branch along with the fundamental solution. Also no transition happens for this solution for the Higgs self-coupling interval of $0\leq \lambda \leq 144$. For the case of $n=3$, two transitions happen along the fundamental solution. Also at a higher energy, there are two bifurcating branches. The lower energy branch of these bifurcating branches, merges with the fundamental solution and both terminate at the convergence point and do not survive for larger values of $\lambda$. For $n=4$, a bifurcation is observed at higher energy in comparison with the fundamental solution. Here there are three transitions. One is observed along the fundamental solution and the others happen along the higher energy bifurcating branch. For the case of $n=5$, the pattern is more complex. A bifurcation in $\lambda = \lambda_{b1}$ happens with a higher energy than the fundamental solution. A second bifurcation is observed at $\lambda = \lambda_{b2}$. The two branches of the second bifurcation are both very close in energy to the lower energy branch of the first bifurcation, but they have different electrical and geometrical properties. Therefore, for the case of $n=5$, we have three distinct solution for the interval of $\lambda_{b1}\leq \lambda \leq \lambda_{b2}$ and five distinct solutions for $\lambda_{b2}\leq \lambda \leq 300$. Also two transitions are observed in the higher energy branch of the first bifurcation.
\end{abstract}

%%-main body of paper-%%
%%-self numbering sections-%%
%%%%%%%%%%%%%%%%%%%%%%%%%%%%%%%%%%%%%************************\section{Introduction}**********************************************************************************

\section{Introduction}
\label{section.1}

Several monopole solutions have been found for SU(2) Yang-Mills-Higgs (YMH) theory which among those some possess both electric and magnetic charges \cite{kn:1}-\cite{kn:4}. The 't Hooft -Polyakov numerical solution with unit topological charge and spherical symmetry, is the first solution of a class of solutions which are invariant under a U(1) subgroup of the local SU(2) gauge group \cite{kn:1}, \cite{kn:2}. This class of solutions  gives rise to Maxwell's electromagnetic field theory after symmetry breaking. The Bogomol'nyi-Prasad-Sommerfield (BPS) limit for which the Higgs potential is zero, is the only condition in which the exact solutions are available \cite{kn:2}, \cite{kn:5}.

Multimonopole solutions in the SU(2) YMH theory with topological charges greater one, cannot be spherically symmetric \cite{kn:6}. A rich class of axially symmetric numerical multimonopole solutions, including monopole-antimonopole pair (MAP), monopole-antimonopole chain (MAC) and vortex-ring configurations are discussed in the ref. \cite{kn:7}. For MAP configurations the Higgs field vanishes at two isolated points along the symmetry axis whereas the number of these isolated points for MAC configurations is more than two. For vortex-rings, the Higgs field vanishes on rings centred around the symmetry axis. A further study by Kunz et al., for $\phi$-winding number of $n=3$ and varying Higgs self-coupling constant, $\lambda$, showed that for the case of two, three and four poles, there are three different branches of solutions with different total energies and geometrical properties \cite{kn:8}. That study indicates that, two of these branches appear with a bifurcation at a critical value of $\lambda$ and a higher energy in comparison with the fundamental solution which appear at $\lambda=0$ . Also the transition between vortex-ring and MAC/MAP configurations was first introduced in that study.

An electrically charged monopole is called a dyon. Axially symmetric dyon solutions with electric charge parameter $0\leq \eta \leq 1$, were first introduced by Hartmann $et~al$. \cite{kn:9}. They showed that for any YMH solution in BPS limit, an electrically charged family of solution can be found. MAP solutions with a critical electric charge were studied in detail in ref. \cite{kn:10} where a one-dipole and a one-vortex-ring configurations were obtained for different values of $\phi$-winding number $n=1,2,3,4$ and 5 and the Higgs self-coupling $\lambda = 0$ and 1. The dependence of physical and geometrical properties of electrically charged MAP configurations in the Higgs self-coupling constant, $\lambda$, for $\phi$-winding number $n=2,3$ and 4 is summarized in ref. \cite{kn:11} for larger values of $\lambda$. 

%stability
Here, we investigate the physical and geometrical properties of electrically charged MAC configurations with three poles (axially symmetric monopole solutions with vanishing magnetic dipole moment), for $\phi$-winding numbers of $n=2,3,4,5$ and varying $\lambda$ and $\eta$. Any solution in this case is composed of a number of monopoles (or vortex-rings). The energy of these bound states, is smaller than the energy of the same number of single poles (or rings) with infinite separation between them. However, this energy is still larger than the lower bound of BPS. Hence, these sort of solutions are static equilibrium states which are not stable in general and are referred as saddle point solutions \cite{kn:12}.

 The study of ref. \cite{kn:8} which has investigated the electrically neutral MAP and MAC configurations of two, three and four poles for the case of $n=3$ and also the study of ref. \cite{kn:11} which investigates the MAP solutions with $n=2,3,4$ and 5, have found only one bifurcation for each of those cases. For the first time in this study, the presence of two bifurcation points (and therefore 5 separate branches) for the case of $n=5$, is found.

Based on our calculations, in the case of $n=2$, the only available solution for the interval of $0\leq \lambda \leq 144$, is fundamental solution and no transition is observed along this solution at this interval. For the case of  $n=3$, the fundamental solution undergoes two transitions at critical points of $\lambda=\lambda_{t1(n=3)}$ and $\lambda=\lambda_{t2(n=3)}$. Also a bifurcation occurs at $\lambda=\lambda_{b(n=3)}$. The lower energy branch (LEB) in this case joins to the fundamental branch at another critical point of $\lambda=\lambda_{j(n=3)}$ where both branches come to the end and do not survive for larger values of $\lambda$. 

 For the case of $n=4$, a transition is observed along the fundamental solution at $\lambda=\lambda_{t1(n=4)}$. Two new branches appear at the bifurcation point at $\lambda=\lambda_{b(n=4)}$ and the higher energy branch (HEB) undergoes transitions at critical points of $\lambda=\lambda_{t2(n=4)}$ and $\lambda=\lambda_{t3(n=4)}$.
 
 Finally for the case of $n=5$, no transition occurs along the fundamental solution however two bifurcations happen at critical points of $\lambda=\lambda_{b1(n=5)}$ and $\lambda=\lambda_{b2(n=5)}$. Two transitions happen along the higher energy branch of the first bifurcation (HEB1) at $\lambda=\lambda_{t1(n=5)}$ and $\lambda=\lambda_{t2(n=5)}$ while the other branches do not include any transition within the studied interval of $\lambda_{b1(n=5)}\leq \lambda \leq 300$.
 
 All of the seven transition points which are detected in this study are geometrically of three major types which we will refer to them as the \textit{type 1} (or the \textit{reverse type 1}), the \textit{type 2} (or the \textit{reverse type 2}) and the \textit{type 3} transitions. This study indicates that an electric and magnetic charge transformation occurs for the pole which is located at the centre during the \textit{type 1} or \textit{reverse type 1} transitions. Also, we have studied the dependence of the position of critical points of $\lambda=\lambda_b, \lambda_t$ and $\lambda_{j}$  with respect to the electric charge parameter $\eta$.

The SU(2) Yang-Mills-Higgs theory and the electromagnetic $Ansatz$ of these new solutions are discussed briefly in the second section. Third section is assigned to numerical procedure and our new results about multiple transitions and bifurcations in the three-poles MAC/vortex-ring configuration and finally we summarize and conclude in the last section.

%%%%%%%%%%%%%%%%%%%%%%%%%%%%%%%The SU(2) YMH Theory%%%%%%%%%%%%%%%%%%%%%%%%%%%%%%%
\section{The SU(2) Yang-Mills-Higgs theory}
\label{sect:2}

The 3+1 dimensional SU(2) YMH Lagrangian is given by

\begin{equation}
{\cal L} = -\frac{1}{4}F^a_{\mu\nu} F^{a\mu\nu} - \frac{1}{2}D^\mu \Phi^a D_\mu \Phi^a - \frac{1}{4}\lambda(\Phi^a\Phi^a - \xi^2)^2, 
\label{eq.1}
\end{equation}

\noindent where the vacuum expectation value of the Higgs field is $\xi=\frac{\mu}{\sqrt{\lambda}}$ in which $\mu$ is the Higgs field mass and $\lambda$ is the Higgs self-coupling constant. The covariant derivative of the Higgs field and the gauge field strength tensor are given respectively by 

\begin{eqnarray}
D_{\mu}\Phi^{a} &=& \partial_{\mu} \Phi^{a} + g\epsilon^{abc} A^{b}_{\mu}\Phi^{c},\nonumber\\
F^a_{\mu\nu} &=& \partial_{\mu}A^a_\nu - \partial_{\nu}A^a_\mu + g\epsilon^{abc}A^b_{\mu}A^c_\nu .
\label{eq.2}
\end{eqnarray}
 
The metric used is $-g_{00}=g_{11}=g_{22}=g_{33}=1$. The SU(2) internal group indices $a, b, c = 1, 2, 3$ and in Minkowski space, $\mu, \nu, \alpha = 0, 1, 2$, and $3$ . The gauge field coupling constant $g$, can be scaled away. Now the Euler-Lagrange equation leads us to the following set of equations of motion 

\begin{eqnarray}
D^{\mu}F^a_{\mu\nu} &=& \partial^{\mu}F^a_{\mu\nu} + \epsilon^{abc}A^{b\mu}F^c_{\mu\nu} = \epsilon^{abc}\Phi^{b}D_{\nu}\Phi^c,\nonumber\\
D^{\mu}D_{\mu}\Phi^a &=& \lambda\Phi^a\left(\Phi^{b}\Phi^{b} - \frac{\mu^2}{\lambda}\right).
\label{eq.3}
\end{eqnarray}
Upon symmetry breaking, the electromagnetic field tensor proposed by 't Hooft is \cite{kn:1} 
\begin{eqnarray}
&&F_{\mu\nu} = \hat{\Phi}^a F^a_{\mu\nu} - \epsilon^{abc}\hat{\Phi}^{a}D_{\mu}\hat{\Phi}^{b}D_{\nu}\hat{\Phi}^c
	= \partial_{\mu}A_\nu - \partial_{\nu}A_\mu - \epsilon^{abc}\hat{\Phi}^{a}\partial_{\mu}\hat{\Phi}^{b}\partial_{\nu}\hat{\Phi}^c, 
\label{eq.4}
\end{eqnarray}

\noindent where, $A_\mu = \hat{\Phi}^{a}A^a_\mu, ~\hat{\Phi}^a = \Phi^a/|\Phi|, ~|\Phi| = \sqrt{\Phi^{a}\Phi^{a}}$. 
We can separate the above mentioned Abelian electromagnetic field into two terms,
\begin{eqnarray}
F_{\mu\nu} = G_{\mu\nu}+H_{\mu\nu}
\label{eq.5}
\end{eqnarray}
\noindent where $ G_{\mu\nu}=\partial_{\mu}A_\nu - \partial_{\nu}A_\mu $ is the Maxwell part and $ H_{\mu\nu}=- \epsilon^{abc}\hat{\Phi}^{a}\partial_{\mu}\hat{\Phi}^{b}\partial_{\nu}\hat{\Phi}^c $ is the Dirac part of the 't Hooft electromagnetic field. For the topological magnetic current density \cite{kn:13} we have
\begin{eqnarray}
k_{\mu} = \frac{1}{8\pi}\epsilon_{\mu\nu\rho\sigma}\epsilon_{abc}\partial^{\nu}\hat{\Phi}^{a}\partial^{\rho}\hat{\Phi}^{b}\partial^{\sigma}\hat{\Phi}^c,
\label{eq.6}
\end{eqnarray}
so that, for the conserved topological magnetic charge carried by the Higgs field, we can write
\begin{eqnarray}
{\cal M} = \int{k_0 d^3x}=\frac{1}{8\pi}\int{\epsilon_{ijk}\epsilon^{abc}\partial_{i}(\hat{\Phi}^{a}\partial_{j}\hat{\Phi}^{b}\partial_{k}\hat{\Phi}^c) d^3x}\nonumber\\ = \frac{1}{8\pi}\oint  \epsilon_{ijk}\epsilon^{abc}\hat{\Phi}^{a}\partial_{j}\hat{\Phi}^{b}\partial_{k}\hat{\Phi}^c~d^{2}\sigma_{i}.
\label{eq.7}
\end{eqnarray}
Furthermore, we know that \cite{kn:14} the topological magnetic charge is the total magnetic charge of the system provided that the gauge field is not singular. In our case in this paper, the gauge field is nonsingular thus we can write the Abelian electric field, $E_i$, the Abelian magnetic field, $B_i$, and the net magnetic charge of the system respectively as below
\begin{eqnarray}
&&E_i =F_{i0}= \partial_{i}A_0 - \partial_{0}A_i,\nonumber\\
&&B_i=-\frac{1}{2}\epsilon_{ijk}F_{jk} = \frac{1}{2}\epsilon_{ijk}\epsilon_{abc}\hat{\Phi}^{a}\partial^{j}\hat{\Phi}^{b}\partial^{k}\hat{\Phi}^c-\frac{1}{2}\epsilon_{ijk}\partial_j A_k, \nonumber\\
&&M = \frac{1}{4\pi} \int \partial^i B_i ~d^{3}x  = \frac{1}{4\pi} \oint d^{2}\sigma_{i}~B_i.
\label{eq.8}
\end{eqnarray}
%%%%%%%%%%%%%%%%%%%%%%%%%%%%%%%%%%The Magnetic Ansatz%%%%%%%%%%%%%%%%%%%%%%%%%%%%%%%%%%%%%%
The $Ansatz$ used for solving axially symmetric dyon solutions is
\begin{eqnarray}
A_i^a &=&  - \frac{1}{r}\psi_1(r, \theta) \hat{n}^{a}_\phi\hat{\theta}_i + \frac{1}{r}\psi_2(r, \theta)\hat{n}^{a}_\theta\hat{\phi}_i
+ \frac{1}{r}R_1(r, \theta)\hat{n}^{a}_\phi\hat{r}_i - \frac{1}{r}R_2(r, \theta)\hat{n}^{a}_r\hat{\phi}_i, \nonumber\\
A^a_0 &=& \tau_1(r, \theta)~\hat{n}^a_r + \tau_2(r, \theta)\hat{n}^a_\theta,
~~\Phi^a = \Phi_1(r, \theta)~\hat{n}^a_r + \Phi_2(r, \theta)\hat{n}^a_\theta.
\label{eq.9}
\end{eqnarray}
Here the spatial unit vectors are given by
\begin{eqnarray}
\hat{r}_i &=& \sin\theta ~\cos \phi ~\delta_{i1} + \sin\theta ~\sin \phi ~\delta_{i2} + \cos\theta~\delta_{i3}, \nonumber\\
\hat{\theta}_i &=& \cos\theta ~\cos \phi ~\delta_{i1} + \cos\theta ~\sin \phi ~\delta_{i2} - \sin\theta ~\delta_{i3}, \nonumber\\
\hat{\phi}_i &=& -\sin \phi ~\delta_{i1} + \cos \phi ~\delta_{i2},
\label{eq.10}
\end{eqnarray}
and the isospin unit vectors are given by
\begin{eqnarray}
\hat{n}_r^a &=& \sin \theta ~\cos n\phi ~\delta_{1}^a + \sin \theta ~\sin n\phi ~\delta_{2}^a + \cos \theta~\delta_{3}^a,\nonumber\\
\hat{n}_\theta^a &=& \cos \theta ~\cos n\phi ~\delta_{1}^a + \cos \theta ~\sin n\phi ~\delta_{2}^a - \sin \theta ~\delta_{3}^a,\nonumber\\
\hat{n}_\phi^a &=& -\sin n\phi ~\delta_{1}^a + \cos n\phi ~\delta_{2}^a.
\label{eq.11}
\end{eqnarray}
The $\phi$-winding number $n$ (which is equal to the net magnetic charge for vanishing magnetic dipole cases) is a natural number. Here, we consider the values of $n=2, 3, 4,$ and 5.
Using the definitions of $h_1(r,\theta) = \Phi_1/|\Phi|$ and $h_2(r,\theta) = \Phi_2/|\Phi|$, the axially symmetric Higgs unit vector will be
\begin{eqnarray}
\hat{\Phi}^a &=& \Phi^a/|\Phi| = h_1(r, \theta)~\hat{n}^a_r + h_2(r, \theta)\hat{n}^a_\theta \nonumber\\
&=& \sin\alpha \cos n\phi ~\delta^{a1} + \sin\alpha \sin n\phi ~\delta^{a2} + \cos\alpha ~\delta^{a3}, 
\label{eq.12}\\
\cos\alpha &=& h_1(r,\theta)\cos\theta - h_2(r,\theta)\sin\theta,\nonumber\\
\sin\alpha &=& h_1(r,\theta)\sin\theta + h_2(r,\theta)\cos\theta.
\label{eq.13}
\end{eqnarray}

%%%%%%%%%%%%%%%%%%%%%%%%%%%%%%%%%%%%%%%%%%%%%%%%%%%%%%%%%%%%%%%%%%%%%%%%%%%%%%%%%%%%%%%%%%%%%%%%%%%%%%%%%%%%%%%%%%%%%%%%%%%%%%%%%%%%%%%%%%%%%%%%%%%%%%%%%%%%%%%%%%%%%

%\subsection{The Magnetic Field and Charge}

Using the eq. (\ref{eq.13}) and the definitions of $\cos\kappa = \frac{\sin\theta}{n}\left(h_2(r,\theta)\psi_2 - h_1(r,\theta)R_2\right)$ and $\gamma=\cos\alpha + \cos\kappa$, the 't Hooft's magnetic field (including both Maxwell part and Dirac part) reduces to
\begin{eqnarray}
B_i &=& -n\epsilon_{ijk}\partial_j\gamma~\partial_k \phi.
\label{eq.14}
\end{eqnarray}

 Based on eq. (\ref{eq.14}), drawing the lines of $\gamma = $ constant, on the vertical plane of $\phi=0$, will represent the magnetic field lines. Also it is easy to see that the unit vectors of magnetic field is given by:
\begin{eqnarray}
\hat{B}_i &=& \frac{r~\partial_{r}(\gamma)\hat{\theta}_i-\partial_{\theta}(\gamma)\hat{r}_i}{\sqrt{(r~\partial_{r}(\gamma))^2+(\partial_{\theta}(\gamma))^2}}.
\label{eq.15}
\end{eqnarray} 

%%%%%%%%%%%%%%%%%%%%%%%%%%%%%%%%%%%%%%%%%%%%%%%%%%%%%%%%%%%%%%%%%%%%%%% %%%%%%%%%%%%%%%%%%%%%%%%%%%%%%%%%%%%%%%%%%%%%%%%%%%%%%%%%%%%%%%%%%%%%%%%%%%%%%%%%%%%%%%%%%%%%%

%\subsection{The Electric Field and Charge}

Since the gauge field is time independent, the Abelian electric field becomes
\begin{eqnarray}
E_i = \partial_i A_0 = \partial_i (\tau_1(r,\theta) h_1(r,\theta) + \tau_2(r,\theta) h_2(r,\theta)).
\label{eq.16}                    
\end{eqnarray}
Now, like the magnetic field, we can construct the unit vectors of the electric field as well.
\begin{eqnarray}
\hat{E}_i &=& \frac{r~\partial_{r}A_0\hat{r}_i+\partial_{\theta}A_0\hat{\theta}_i}{\sqrt{(r~\partial_{r}A_0)^2+(\partial_{\theta}A_0)^2}}.
\label{eq.17}
\end{eqnarray} 

 At spatial infinity in the Higgs vacuum, the time component of the gauge field is parallel to Higgs field in isospin space \cite{kn:9} , \cite{kn:10} and the proportionality constant is the electric charge parameter, $\eta$. Then at large distances we can write:
\begin{eqnarray}
E_i = \partial_i |\tau|=\partial_i\sqrt{\tau_1^2+\tau_2^2}.
\label{eq.18}                    
\end{eqnarray}
Therefore, the electric field varies proportionally with the electric charge parameter, $0\leq\eta< 1$ and then can be switched off by setting $\eta=0$. 
The contour plot of the time component of the gauge potential, $A_0=$ constant, gives the equipotential lines of the electric field which are always perpendicular to the electric field vectors. Also the total electric charge of the system, $Q$ can be evaluated numerically by 
\begin{equation}
Q=\frac{1}{4\pi\xi}\int{\partial^iE_i}~d^3x.
\label{eq.19}                    
\end{equation}

%%%%%%%%%%%%%%%%%%%%%%%%%%%%%%%%%%%%%%%%%%%%%%%%%%%%%%%%%%%%%%%%%%%%%%%%%%%%%%%%%%%%%%%%%%%%%%%%%%%%%%%%%%%%%%%%%%%%%%%%%%%%%%%%%%%%%%%%%%%%%%%%%%%%%%%%%%%%%%%%%%%%%

%\subsection{The Magnetic Dipole Moment and the Angular Momentum}
From Maxwell's electromagnetic theory, the dimensionless magnetic dipole moment for axially symmetric MAP/MAC solutions, is given by $\mu_m=-\frac{F_G(\theta)}{\sin^2\theta}$, where $F_G$ is given by \cite{kn:10}:
\begin{equation}
F_G(\theta)=r\sin\theta\left\{h_1(n\cot\theta-R_2)+h_2(\psi_2-n)-\frac{n}{\sin\theta}(a\cos\theta+b)\right\}|_{r\rightarrow\infty}, 
\label{eq.20}
\end{equation}
with $a=0,~b=1$ for configurations with an even number of poles and $a=1,~b=0$ for the case of an odd number of poles.
In the MAC system of solutions with odd number of poles, the symmetry of magnetic charge with respect to the $x$-$y$ plane (the $z$ axis is the symmetry axis), causes the magnetic dipole moment to vanish.

The angular momentum density is defined by \cite{kn:10}:

\begin{eqnarray}
j_z&=&\epsilon_{kij}~\hat{\rho_i}~\rho~ \theta_{0j}~ \delta^3_k,\nonumber\\
\theta_{0j}&=&F^{ai}_0 F^a_{ij}+D_0\Phi^a D_j\Phi^a.
\label{eq.21}                    
\end{eqnarray}

Using the same convention of eq. (\ref{eq.20}) for $a$ and $b$ for axially symmetric MAP/MAC systems we have \cite{kn:10}, \cite{kn:15}
\begin{eqnarray}
J_z=\frac{n}{2\xi}\int^\pi_0\{r^2\sin\theta~\partial_r |\tau| ~(a\cos\theta+b)\}|_{r\rightarrow\infty}~d\theta~~\Rightarrow\nonumber\\
\end{eqnarray}                    
\[J_z = \left\{
  \begin{array}{ll}
    0 & \quad(a=1,~b=0)\\
    nQ & \quad(a=0,~b=1).
    \label{eq.22}
  \end{array}
\right.
\]
So that for the MAC solutions with an odd number of poles, the angular momentum vanishes as well.

%%%%%%%%%%%%%%%%%%%%%%%%%%%%%%%%%%%%%%%%%%%%%%%%%%%%%%%%%%%%%%%%%%%%%%%%%%%%%%%%%%%%%%%%%%%%%%%%%%%%%%%%%%%%%%%%%%%%%%%%%%%%%%%%%%%%%%%%%%%%%%%%%%%%%%%%%%%%%%%%%%%%%

%\subsection{The Energy}

The electrically charged BPS case defines a lower bound for static energy which is given by  \cite{kn:13}
\begin{equation}
E_{min} = 4\pi\xi\sqrt{{\cal M}^2 + Q^2}. 
\label{eq.23}
\end{equation}
Hence, the dimensionless total energy of the MAC dyon solution even in the limit of vanishing $\lambda$, is more than this lower bound and is given by \cite{kn:7}, \cite{kn:14},
\begin{eqnarray}
E=\frac{1}{4\xi}\int{\left\{B^a_iB^a_i + E^a_iE^a_i + D_i\Phi^aD_i\Phi^a + D_0\Phi^aD_0\Phi^a + \frac{\lambda}{2}(\Phi^a\Phi^a-\xi^2)^2\right\}d^2x}.
\label{eq.24}
\end{eqnarray}

%%%%%%%%%%%%%%%%%%%%%%%%%%%%%%%%%%%%%%%%%%%%%%%%%%%%%%%%%%%%%%%%%%%%%%%%%%%%%%%%%%%%%%%%%%%%%%%%%%%%%%%%%%%%%%%%%%%%%%%%%%%%%%
\section{The numerical solution}
\label{sect:3}

\subsection{The boundary conditions}
\label{sect:3.1}
Equations of motion (\ref{eq.3}) with the gauge field and Higgs field of $Ansatz$ (\ref{eq.9}), lead us to a system of eight coupled nonlinear second order equations for eight profile functions of the  electromagnetic $Ansatz$. The boundary conditions at large distances are given by \cite{kn:7}, \cite{kn:16}:

\begin{eqnarray}
&&\psi_1(r,\theta)|_{r\rightarrow\infty} = 3, ~~\psi_2(r,\theta)|_{r\rightarrow\infty} = \frac{n(\sin\theta+\cos\theta~\sin 2\theta)}{\sin\theta}=n(\cos2\theta+2),\nonumber\\
&&R_1(r,\theta)|_{r\rightarrow\infty} = 0,~~ R_2(r,\theta)|_{r\rightarrow\infty} =\frac{n(\cos\theta-\cos\theta~\cos 2\theta)}{\sin\theta}=n\sin2\theta,\nonumber\\
&&\Phi_1(r,\theta)|_{r\rightarrow\infty} = \xi \cos 2\theta, ~\Phi_2(r,\theta)|_{r\rightarrow\infty} = \xi \sin 2\theta, ~\nonumber\\
&&\tau_1(r,\theta)|_{r\rightarrow\infty} = \eta~ \xi\cos 2\theta, ~\tau_2(r,\theta)|_{r\rightarrow\infty} = \eta~ \xi\sin 2\theta.
\label{eq.25}
\end{eqnarray}
 As we already mentioned (and is formulated in the condition above), the Higgs field and the time component of the gauge are supposed to be parallel in isospin space at large distances. The trivial boundary conditions at $r=0$ are given by \cite{kn:7}, \cite{kn:10}: 

\begin{eqnarray}
&&\psi_1=\psi_2=R_1=R_2=0, 
\label{eq.23}\\\nonumber\\
&&\sin\theta~\tau_1(0,\theta)+\cos\theta~\tau_2(0,\theta)=0, ~\sin\theta~\Phi_1(0,\theta)+\cos\theta~\Phi_2(0,\theta)=0,\nonumber\\
&&\partial_r(\cos\theta~\tau_1(r,\theta)-\sin\theta~\tau_2(r,\theta))|_{r=0}=0,\nonumber\\
&&\partial_r(\cos\theta~\Phi_1(r,\theta)-\sin\theta~\Phi_2(r,\theta))|_{r=0}=0.
\label{eq.26}
\end{eqnarray}

and along the $z$-axis we have \cite{kn:7}, \cite{kn:10}:
\begin{eqnarray}
R_A(r, \theta)|_{\theta \rightarrow 0, ~\pi} = \Phi_2(r, \theta)|_{\theta \rightarrow 0, ~\pi} = \tau_2(r, \theta)|_{\theta \rightarrow 0, ~\pi} &=& 0, \nonumber\\
\partial_{\theta}\psi_A(r, \theta)|_{\theta \rightarrow 0, ~\pi} = \partial_{\theta}\Phi_1(r, \theta)|_{\theta \rightarrow 0, ~\pi} = \partial_{\theta}\tau_1(r, \theta)|_{\theta \rightarrow 0, ~\pi} &=& 0,
\label{eq.27}
\end{eqnarray}
where $A=1, 2$. Here we choose the value of $\xi$ to be one.  These sets of conditions together with gauge fixing condition $r\partial_rR_1-\partial_\theta \psi_1=0$, \cite{kn:7} gives  us the complete set of boundary conditions of MAC dyon solutions with three poles.

\subsection{The numerical method}
\label{sect:3.2}
The numerical procedure for this problem consists of two major steps. The first step is implemented in Maple and  the second one is executed in MATLAB. In the first step, finite difference method is used to provide approximations for terms including derivatives in equations of motion. Also Maple is used to produce a Jacobian sparsity pattern for the system of equations. In the second step, the trust-region-reflective algorithm is used to solve the system of nonlinear partial differential equations by finding the roots of linear approximations of the nonlinear system (linearization).

The trust-region optimization methods are typically very sensitive to initial approximation \cite{kn:17},\cite{kn:18}. This means that the quality of convergence \footnote{The quality of convergence is given by two parameters. Firstly, how small is the main function $f(x)$ which is supposed to become minimized and secondly, how small is first order optimality which is a measure to show how close is the point $x$ to optimal \cite{kn:19}.} and consequently the accuracy of our final result for total energy, total electric charge and geometrical configuration of the charge distribution, are quite sensitive to our initial guess for optimized function.  A good choice for the initial guess to find a solution for $\lambda_i+\delta \lambda$, is the solution which is already obtained for $\lambda_i$.  Generally, a smaller value for $\delta \lambda$, causes a better convergence. This means that in order to get a more accurate solution, we need to choose smaller steps and consequently higher number of optimization processes.

On the other hand, according to our previous experiences of numerical calculations in Ref \cite{kn:11}, the stopping criteria of the fsolve package in MATLAB are not adequately accurate in some cases and we need to run the optimization toolbox for a larger number of iterations in order to obtain smoother diagrams for the solutions. Both of these limitations make it necessary to run the optimization toolbox for thousands of times to get a clear and accurate picture of final solution.

In our previous study of ref. \cite{kn:11}, we used a version of numerical method in which all the numerical data processing steps were manually done. But to avoid the above mentioned inaccuracy, we need to generate a huge amount of numerical data for which the manual processing method will not be adequate.

For the current study, we generated a new version of the numerical method in which the numerical data processing steps are controlled automatically. Also this new method confines the first order optimality of the solutions to a reasonable amount and controls the necessary number of iterations for a proper convergence.

Our polar grid of the size $70\times60$ covers the region of $0\leq \bar{x}=\frac{r}{r+1}\leq 1$, and $0\leq \theta \leq \pi$. There are two major kinds of errors in our numerical solutions. One of those come from the finite difference approximation. In finite difference method, the error depends on the seleted type of approximation and the size of the steps of the grid. The error of our central difference approximation is of the order of $\left(\frac{\pi}{2\times 60}\right)^2\approx 6.85\times 10^{-4}$. The origin of the error of the second kind is the trust-region-reflective optimization method and this error is quite dependent on the quality of the convergence. The vector of the functions which is supposed to be zero, after the convergence still has a small non zero value which is usually of the order of $10^{-6}$. It is clear that the effect of the first kind of the errors will be dominant and therefore the errors in our solutions are of the order of $10^{-4}$.\footnote{Near the bifurcation points, the quality of convergence decreases and the error of the linearization process some times is of the order of $10^{-3}$. Therefore higher orders of error are possible for those solutions.}

%%%%%%%%%%%%%%%%%%%%%%%%%%%%%%%%%%%%%%%%%%%%%%%%%%%%%%%%%%%%%%%%%%%%%%%%%%%%%%%%%%%
%%%%%%%%%%%%%%%%%%%%%%%%%%%%%%%%The Numerical Results%%%%%%%%%%%%%%%%%%%%%%%%%%%%%%%%%%%%
\subsection{The numerical results}
\label{sect:3.3}

This study investigates the numerical solutions for the cases with the $\phi$-winding number $n=2, 3, 4$, and 5 and the electric charge parameter, $0\leq\eta<1$. The interval probed for the Higgs coupling constant\footnote{For all of the cases we have investigated larger intervals however the results are presented for those above mentioned intervals in which the valuable data is available.} for $n=2, 3$ and 4 is $0\leq\lambda\leq 144$ and for $n=5$ is $0\leq\lambda\leq 300$. 
The fundamental solutions in all cases possess the smallest value of energy. In case of $n=2$, no bifurcation (new branching solution) is found. In case of $n=3$ there are four critical points including two transition points, a bifurcation point and a joining point. For the case of $n=4$, three transitions and one bifurcation are detected and finally in the case of $n=5$, there are two bifurcations, and two transitions. The energies of the branches of the second bifurcation of $n=5$, are very close to the energy of the lower energy branch of the first bifurcation but geometrical studies beside their electric charge show that these solutions are different solutions with near energies. 

All of the solutions mentioned in this paper possess the positive net magnetic charge of $n$. Transitions in some cases cause the charge distribution to change but the total magnetic charge of the system is always equal to the $\phi$-winding number of the system.

We will refer to the distance of the vortex-rings from $x$-$y$ plane as $D_z$. The distance of the magnetic poles from $x$-$y$ plane is shown by $d_z$ and the diameter of vortex-rings is shown by  $D_\rho$. 
\subsubsection{The $n=2$ case}
\label{sect:3.3.1}

\begin{figure}[tbh]
	\hskip-0.2in
	\includegraphics[width=6.3in,height=2in]{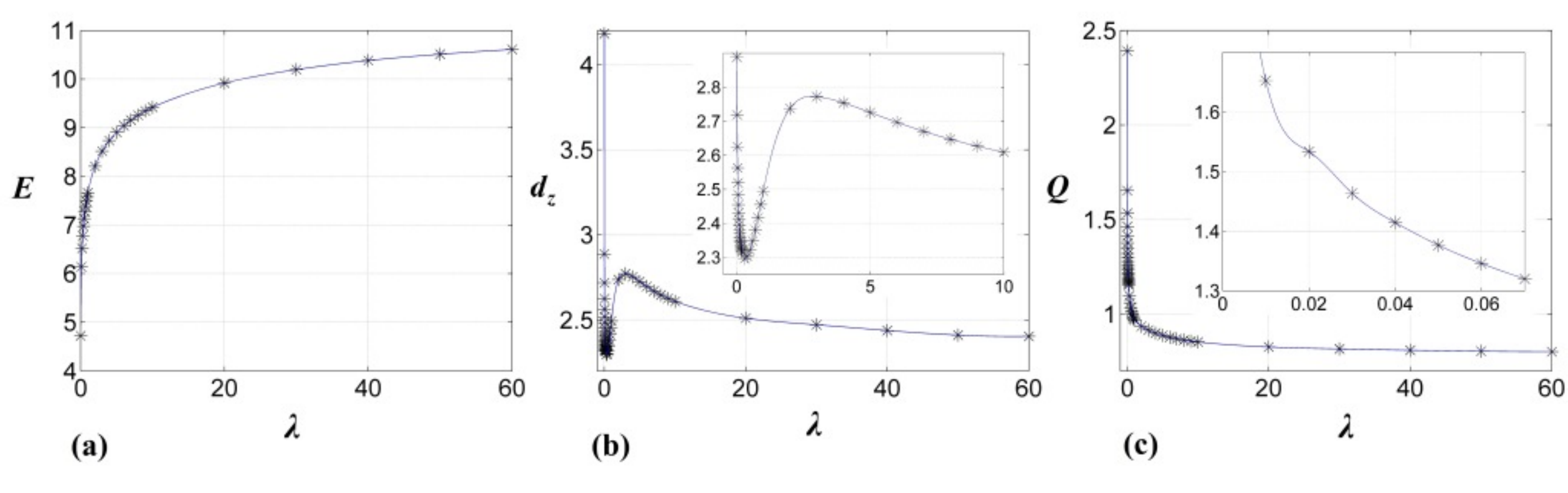} %hskip-0.0in
	\caption{Plots of (a) the total energy, $E$, (b) the distance of the poles from the centre, $d_z$, and (c) the total electric charge, $Q$,  versus the Higgs self-coupling, $\lambda$, when $n=2$, $\eta=0.5$. }
\label{fig.1}
\end{figure}

This case is a simple case including only the fundamental solution which does not undergo any transition. This solution keeps the three-poles form for the interval of study, $0\leq \lambda\leq 144$. As is illustrated in figure \ref{fig.1} the separation between poles with unlike magnetic charges, has a minimum value when $\lambda$=0.3192 and a local maximum value when $\lambda$=2.8021. The electric charge of the solution and the separation of the poles experience a fast drop for $0 \leq \lambda < 0.1$, while the total energy increases rapidly within this interval. The detailed information about the total energy, electric charge and distance of poles from the centre are summarized in table \ref{table.1}.

\begin{table}[tbh]
\small
\hskip-0.4in
\begin{tabular}{|c|cccccccccc|}

\hline
  \multicolumn{11}{|c|} {Fundamental Solution ~~($n=2, ~\eta=0.5$) }  \\ 
\hline

$\lambda$ & 0&	0.01&	0.1&	1&	10&	20&	50&	80&	100&	144\\ 
\hline
$E$ & 4.7219&	5.3082&	6.1378&	7.6607&	9.4225&	9.9189&	10.5123&	10.7535&	10.8515&	10.9841\\ 
\hline
$d_z$ & 4.1821&	2.8877&	2.3952&	2.4945&	2.6091&	2.5105&	2.4112&	2.4039&	2.4065&	2.4103\\
\hline
$Q$ & 2.3907&	1.6526&	1.2616&	0.9695&	0.8527&	0.8265&	0.8040&	0.7991&	0.7980&	0.7966\\ 
\hline

\end{tabular}
\caption{\label{table.1} Table of the dimensionless total energy $E$, the poles' separation $d_z$,  and the electric charge $Q$, of the fundamental solution, when $n=2$, $\eta=0.5$.}
\end{table}

\subsubsection{The $n=3$ case}
\label{sect:3.3.2}

Kunz et al. \cite{kn:8} have studied the configurations of this case and the related transitions for electrically neutral case and a smaller interval of Higgs self-coupling constant, $\lambda$. Here the case is studied in presence of electric charge and larger values of $\lambda$. The configuration of magnetic charge for the fundamental solution in this case includes two vortex-rings which are symmetric with respect to the origin and a positively charged magnetic monopole at the centre for $\lambda=0$ (figure \ref{fig.5}d).  As $\lambda$ increases the diameter and separation of vortex-rings decrease. At a critical value of $\lambda=\lambda_{t1(n=3)}$, in a transition, two new poles emerge from the centre (figure \ref{fig.5}c). The value of $\lambda$ for this critical point for $\eta=0.5$ is $\lambda_{t1(n=3)}=2.557$. We call this kind of transition as \textit{type 1}. These new poles move further away from each other along the symmetry axis (figure \ref{fig.5}b). In another critical point with $\lambda=\lambda_{t2(n=3)}$, the vortex-rings merge with these new poles on the symmetry axis and the configuration changes into a three-poles MAC configurations (figure \ref{fig.5}a). The value of $\lambda$ for this critical point for $\eta=0.5$ is $\lambda_{t2(n=3)}=3.079$. This second transition is called a \textit{type 2} transition in this article.\footnote{During the transitions of \textit{type 2}, the poles are always surrounded with very small rings.} So that, for $\lambda<\lambda_{t1(n=3)}$, we have two vortex-rings and a pole. For $\lambda_{t1(n=3)}\leq\lambda<\lambda_{t2(n=3)}$, there are two vortex-rings and three poles and for $\lambda > \lambda_{t2(n=3)}$, we have three poles. During the transition point of $\lambda=\lambda_{t1(n=3)}$, the magnetic charge of the pole which is located at centre changes from positive to negative charge.

\begin{figure}[tbh]
\centering
 \includegraphics[width=3in,height=2.5in]{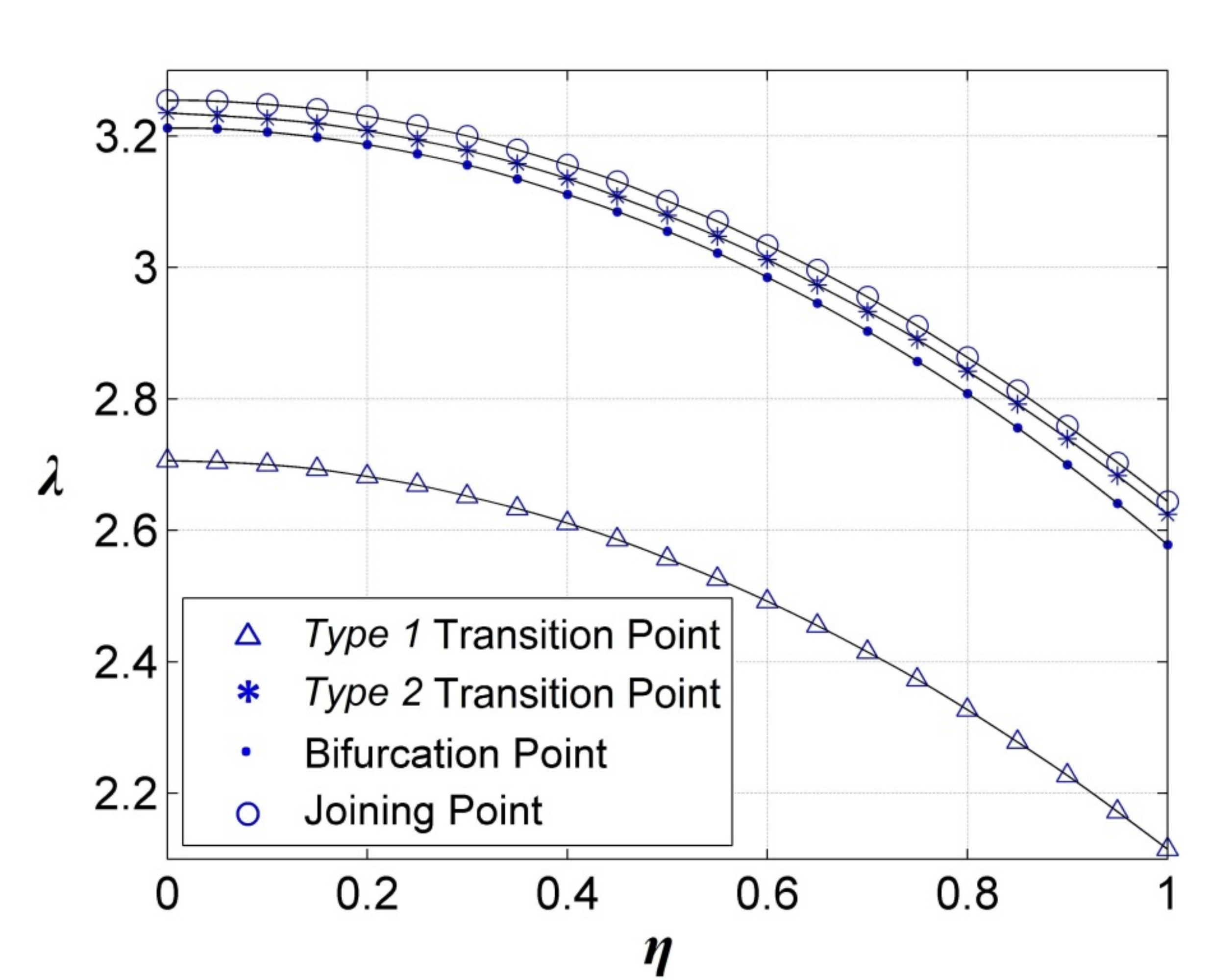} %\hskip-0.0in
	\caption{Higgs self-coupling constant $\lambda$, for transition, bifurcation and joining points versus the electric charge parameter $\eta$, for the case of $n=3$.}
	\label{fig.2}
\end{figure}

Beside the fundamental solution and at a higher energy, two new branches of solution appear in a bifurcation point with $\lambda=\lambda_{b(n=3)}$. The value of $\lambda$ for this critical point for $\eta=0.5$ is $\lambda_{b(n=3)}=3.055$. The configuration of both new branches is the three-poles MAC configuration. The higher energy branch (HEB) survives for $\lambda_{b(n=3)}\leq\lambda\leq 144$, and no transition happens in its configuration but the lower energy branch (LEB) survives only within the small interval of $\lambda_{b(n=3)}\leq\lambda\leq \lambda_{j(n=3)}$, and at the critical point of $\lambda=\lambda_{j(n=3)}$ joins to the fundamental solution and both solutions stop at this point and do not survive for larger values of $\lambda$. The value of $\lambda$ for the joining point for $\eta=0.5$ is $\lambda_{j(n=3)}=3.101$. So that for the intervals of $0\leq\lambda\leq \lambda_{b(n=3)}$ and $\lambda_{j(n=3)}\leq\lambda\leq 144$, there is just one solution while for the interval of $\lambda_{b(n=3)}\leq\lambda\leq \lambda_{j(n=3)}$, we have all three solutions including fundamental, LEB and HEB solutions. Table \ref{table.2} summarizes the values of $\lambda$ in which the transitions, the bifurcation and the joining of branches happen for different values of $\eta$. The general sequence of the location of the critical points, $\lambda_{t1(n=3)}<\lambda_{b(n=3)}<\lambda_{t2(n=3)}<\lambda_{j(n=3)}$, is valid for all values of electric charge parameter, $\eta$, as shown in figure \ref{fig.2}.

\begin{figure}[tbh]
 \includegraphics[width=6in,height=4.1in]{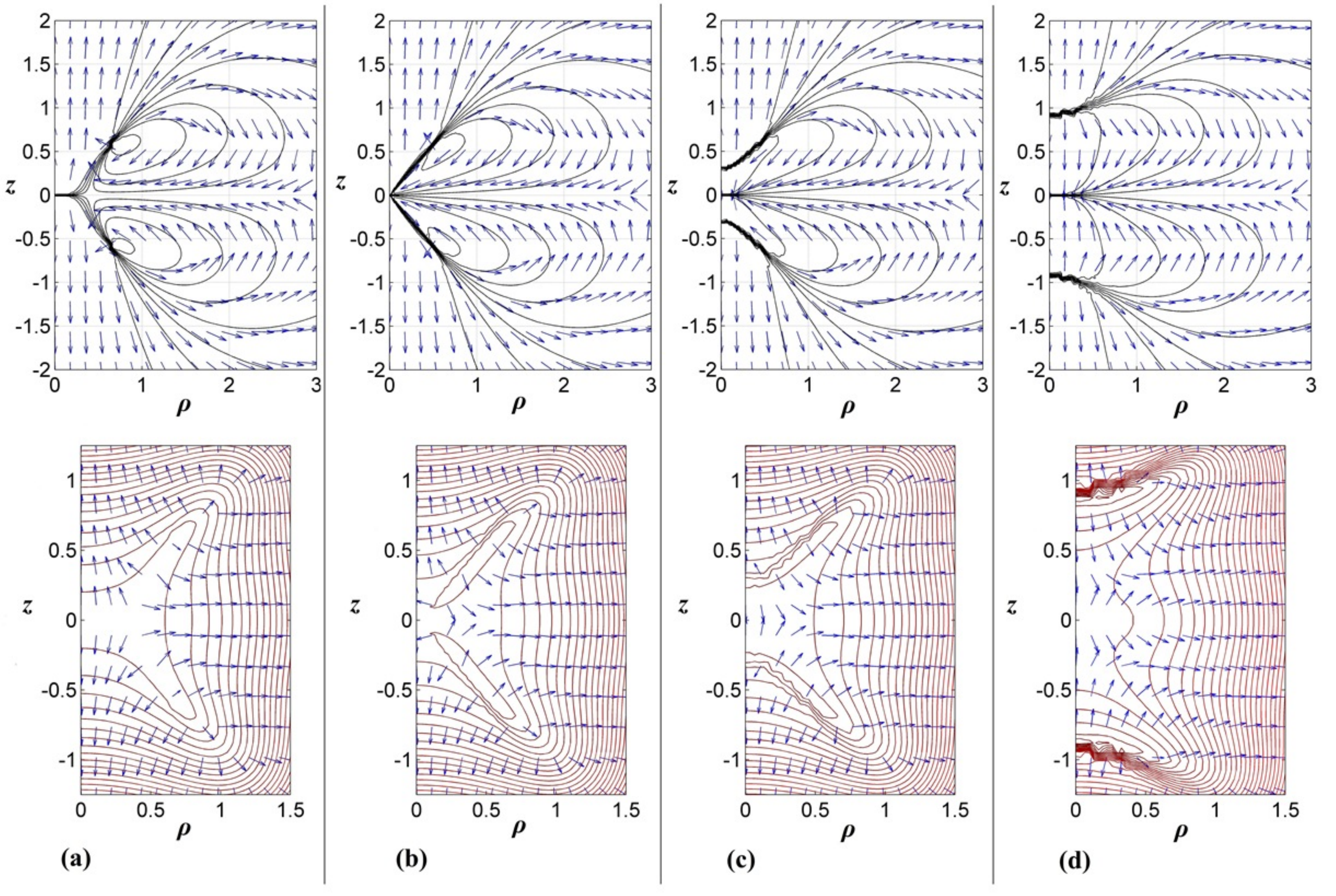} %\hskip-0.0in
	\caption{Magnetic field lines and magnetic field's unit vectors (top) and equipotential lines and unit vectors of electric field (bottom) of the fundamental solution for the case of $n=3$, $\eta=0.5$ when, (a) $\lambda=2$, where we have a pole and two rings, (b) $\lambda = \lambda_{t1(n=3)}=2.557$, where the transition of \textit{type 1} occurs, (c) $\lambda=2.7$, where we have three poles and two rings and (d) $\lambda=3.1$, where we have three poles after going through a \textit{type 2} transition.}
	\label{fig.3}
\end{figure}

For the electrically neutral case of $\eta=0$, these values are $\lambda_{t1(n=3)}=2.706$, $\lambda_{b(n=3)}=3.212$, $\lambda_{t2(n=3)}=3.235$ and $\lambda_{j(n=3)}=3.254$. \footnote{Kunz et al. \cite{kn:8} have investigated the electrically neutral case and their values for critical points are $\lambda^k_{t1(n=3)}=0.673$, $\lambda^k_{b(n=3)}=0.807$, $\lambda^k_{t2(n=3)}=0.810$ and $\lambda^k_{j(n=3)}=0.819$. Regarding to the fact that $\lambda=4\lambda^k$, our results are very close to their results.}  

\begin{figure}[tbp]
	\includegraphics[width=6in,height=5in]{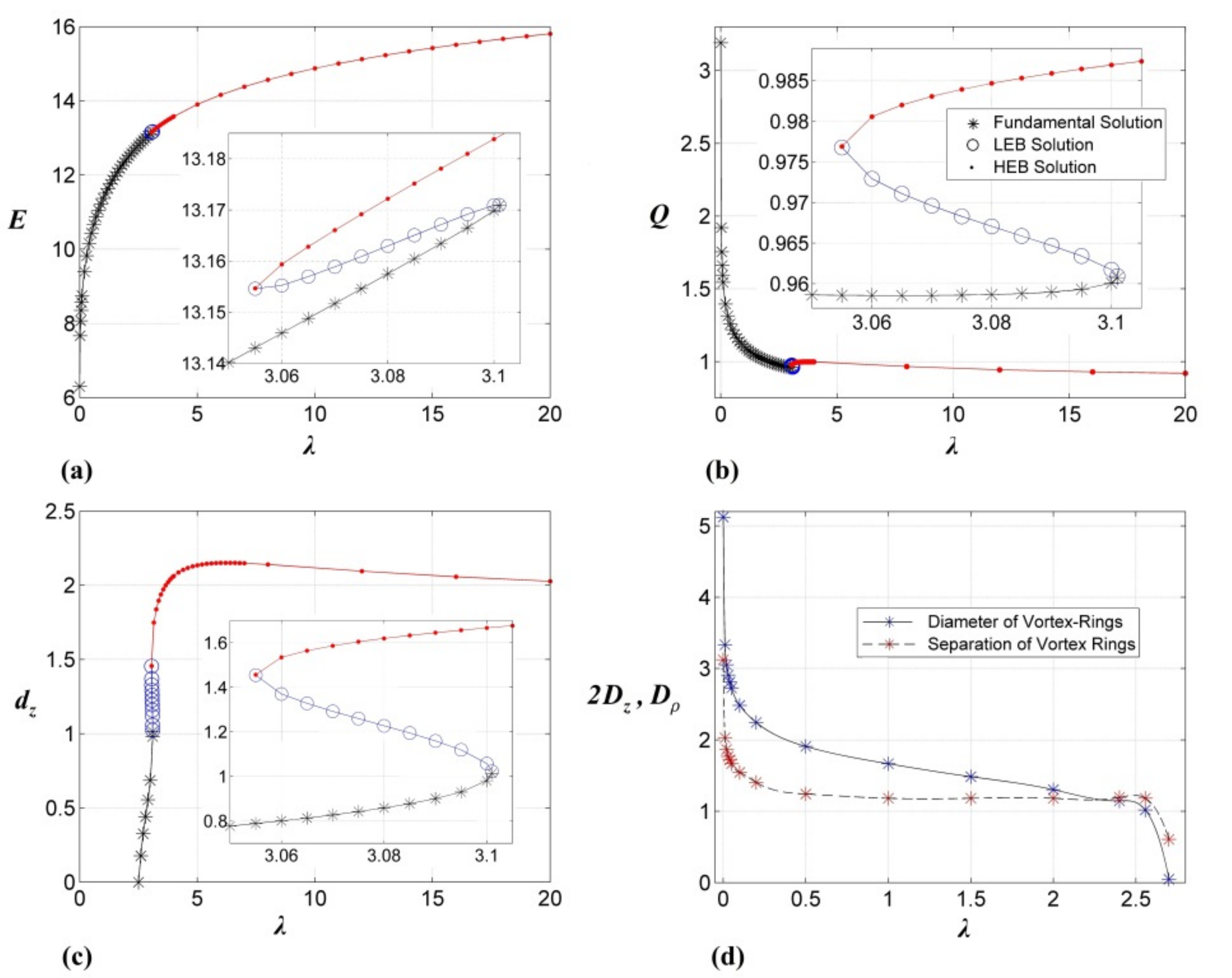} %\hskip-0.0in
	\caption{Plots of (a) the total energy, $E$, (b) the total electric charge, $Q$, (c) the distance of the poles from the centre, $d_z$, and (d) the separation of vortex-rings, $2D_z$, and diameter of vortex-rings, $D_\rho$, versus the Higgs self-coupling, $\lambda$, when $n=3$, $\eta=0.5$. }
	\label{fig.4}
\end{figure}
Integration on small volume including the origin for the fundamental solution, shows that the pole which is located at the centre has a very small positive electric charge. Surprisingly, for the small interval of $\lambda_{t1(n=3)}<\lambda<\lambda_{t2(n=3)}$, electric charge of the central pole becomes negative in sign but very small in magnitude (practically neutral).  However, again for the interval of  $\lambda>\lambda_{t2(n=3)}$, this pole acquires positive electric charge. This positive charge is very much smaller in comparison with that for the other poles which are located symmetrically on the z-axis. The LEB and the HEB solutions for the case of $n=3$, evidently have central poles with positive electric charges.

Figure \ref{fig.3} which shows the behaviour of magnetic and electric fields, is an illustration of the steps of the transitions for $\eta=0.5$. The change in the sign of electric and magnetic charge of the pole which is located at the centre can be seen in this figure as well.
%%%%%%%%%%%%%%%%%%%%%%%%%%%%%%%%%%%%%%%%%%%%%%%%%%%%%%%%%%%%%%%%%%%%%%%%%%%%%%%%%%%%%%%%%%%%%%%%%%

\begin{table}[tbh]
\small
\hskip-0.5in
\begin{tabular}{|c|ccccccccccc|}
\hline
  \multicolumn{12}{|c|} {Critical Points for the Case of~$n=3$}  \\ 
\hline

$\eta$ & 0&	0.1&	0.2&	0.3&	0.4&	0.5&	0.6&	0.7&	0.8&	0.9& 1\\ 
\hline
$\lambda$~(\textit{Type 1} Trans.) & 2.706&	2.700&	2.682&	2.652&	2.611&	2.557&	2.492&	2.415&	2.327&	2.227&	2.115\\ 
\hline
$\lambda$~(Bifurcation)  & 3.212&	3.206&	3.187&	3.156&	3.111&	3.055&	2.985&	2.903&	2.808&	2.700&	2.578\\
\hline
$\lambda$~(\textit{Type 2} Trans.) & 3.235&	3.226&	3.208&	3.178&	3.135&	3.079&	3.012&	2.933&	2.842&	2.739&	2.624\\ 
\hline
$\lambda$~(Joining~Point) & 3.254&	3.248&	3.230&	3.200&	3.156&	3.101&	3.034&	2.955&	2.863&	2.759&	2.644\\
\hline

\end{tabular}
\caption{\label{table.2} Table of the critical values of $\lambda$ for which the transitions of \textit{type 1} and \textit{type 2} and bifurcation and joining of branches happen, for $n=3$.}
\end{table}

Figure \ref{fig.4} illustrates the total energy, the total electric charge and the geometrical parameters of the solutions versus Higgs self-coupling constant $\lambda$. As can be seen in this figure, for the case of $\eta=0.5$, the electric charge of the fundamental solution has a minimum in $\lambda=3.065$. The electric charge of the HEB solution has a maximum at $\lambda=3.7194$ and finally the distance of the poles from the centre of HEB solution has a maximum at $\lambda=6.2843$. More detailed quantitative data of the case $n=3$ are summarized in table \ref{table.3}.

\begin{table}[tbh]
\small
\hskip-0.5in
\begin{tabular}{|c|cccccccccc|}
\hline
  \multicolumn{11}{|c|} {Fundamental Solution ~~($n=3, ~\eta=0.5$) }  \\ 
\hline
 & \multicolumn{3}{c|}{1~Pole~and~2~Rings}&\multicolumn{4}{c|}{3~Poles~and~2~Rings}&\multicolumn{3}{c|}{3~Poles}\\
 \hline
$\lambda$ &0 &	0.1&\multicolumn{1}{c|} {1}	&2.557	&	3&	3.06&	\multicolumn{1}{c|} {3.07}&	3.08&	3.09&	3.1\\ 
\hline
$E$ & 6.3027&	8.7484&	\multicolumn{1}{c|} {11.3685}&	12.8398&	13.1114&	13.1459&\multicolumn{1}{c|} {13.1517}	&	13.1575&	13.1634&	13.1698\\ 
\hline
$d_z$ & -&-&\multicolumn{1}{c|} {-}&0.0146&0.6860	&	0.8005&\multicolumn{1}{c|} {0.8267}	&	0.8585&	0.9005&	0.9808\\
\hline
$D_z$ & 1.5610&	0.7727&	\multicolumn{1}{c|} {0.5909}&0.5909	&overlap&overlap&\multicolumn{1}{c|} {overlap}&-&-&-\\
\hline
$D_\rho$ & 5.1220&	2.4840&	\multicolumn{1}{c|} {1.6666}&1.0176	&$<0.05$&$<0.05$&\multicolumn{1}{c|} {$<0.05$}&-&-&-\\
\hline
$Q$ & 3.1907&	1.5443&	\multicolumn{1}{c|} {1.1028}&0.9752	&	0.9594&	0.9585&	\multicolumn{1}{c|} {0.9586}&	0.9587&	0.959&	0.9601\\ 
\hline

\multicolumn{11}{|c|} {3~Poles~LEB ~~($n=3, ~\eta=0.5$) }  \\ 
\hline
$\lambda$ &&&&&& 3.06&	3.07&	3.08&	3.09&	3.1\\ 
\hline
$E$ &&&&&& 13.1552&	13.1589&	13.163&	13.1671&	13.1709\\ 
\hline
$d_z$ &&&&&& 1.3702&	1.2919&	1.2269&	1.1593&	1.0580\\
\hline
$Q$ &&&&&& 0.9729&	0.9696&	0.9671&	0.9647&	0.9617\\

\hline
\multicolumn{11}{|c|} {3~Poles~HEB ~~($n=3, ~\eta=0.5$) }  \\ 
\hline

$\lambda$ & 3.06&	3.07&	3.08&	3.09&	3.1&	4	&10&	40&	100&	144\\ 
\hline
$E$ & 13.1593&	13.1660&	13.1722&	13.1781&	13.1838&	13.5798&	14.8735&	16.6732&	17.6556&	17.9519\\ 
\hline
$d_z$ & 1.5346&	1.5869&	1.6199&	1.6455&	1.6671&	2.0628&	2.118&	1.9564&	1.9530&	1.9669\\
\hline
$Q$ & 0.9806&	0.9831&	0.9847&	0.9859&	0.9869&	0.9981&	0.954&	0.8953&	0.8816&	0.8793\\
\hline

\end{tabular}
\caption{\label{table.3} Table of the dimensionless total energy $E$, the poles' separation $d_{z}$, the diameter of vortex-rings $D_{\rho}$, the distance of vortex-rings from $x$-$y$ plane $D_z$, and the electric charge $Q$, of different solutions, when $n=3$, $\eta=0.5$. The size of our grid does not let us to calculate the accurate separation and diameter of very small vortex-rings (for $2.7<\lambda<3.079$, $D_\rho<0.05$) because the poles and the vortex-rings are very cloes to each other and their fields overlap in the small area around the poles. Then the value of $D_{z}$ obviously has to be very cloes to the value of $d_{z}$ for this interval.}
\end{table}

\begin{figure}[tbh]
	\includegraphics[width=6in,height=7in]{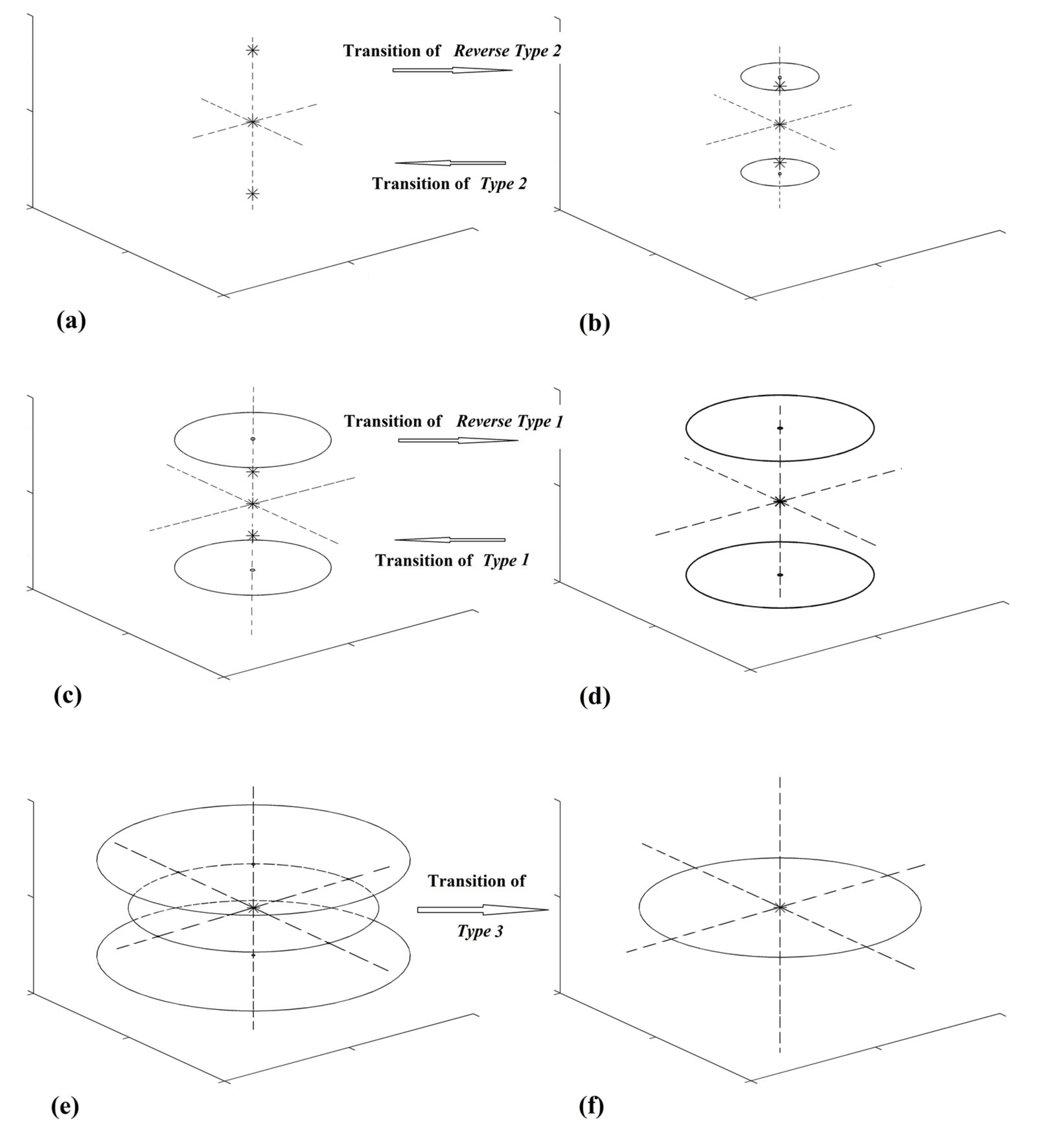} %\hskip-0.0in
	\caption{Schematic illustration of three major transitions. For the \textit{reverse type 2} transition, in the beginning (a) there are three poles on the symmetry axis and then (b) two vortex-rings emerge from the two outer poles. Before the \textit{reverse type 1} transition (c) the separation of the poles decrease and at the transition point, (d) the poles merge to each other on the $x$-$y$ plane. For the transition of \textit{type 3}, the configuration of (e) three vortex-rings and one pole at the centre changes to (f) a vortex-ring on $x$-$y$ plane and a pole at the centre.}
	\label{fig.5}
\end{figure}

\subsubsection{The $n=4$ case}
\label{sect:3.3.3}

For this case, the fundamental solution starts with two symmetric vortex-rings with respect to the origin, a third vortex-ring with smaller diameter on $x$-$y$ plane and a pole at the centre (figure \ref{fig.5}e).  As $\lambda$ increases, the diameter of symmetric vortex-rings and their separation decrease and finally they merge with the third vortex-ring at the transition point of $\lambda=\lambda_{t1(n=4)}$ (figure \ref{fig.5}f). We call this new kind of transition as a \textit{type 3} transition. The two different configurations of fundamental solution for electrically neutral case and small values of $\lambda$ are studied by Kleihaus et al. \cite{kn:7}. The transition of \textit{type 3} for $\eta=0.5$ occurs at $\lambda_{t1(n=4)}=2.67$. The fundamental branch does not undergo any other transition for the interval of $\lambda_{t1(n=4)}<\lambda\leq 144$.

At higher energy, a bifurcation occurs and two new branches of solution appear at critical point of $\lambda=\lambda_{b(n=4)}$. The bifurcation for $\eta=0.5$, occurs at $\lambda_{b(n=4)}=5.979$. The LEB solution here is a three-poles solution for the interval of $\lambda_{b(n=4)}<\lambda\leq 144$ and does not experience any transition. The HEB solution however, undergoes a \textit{reverse type 2} transition at critical point of $\lambda=\lambda_{t2(n=4)}$, in which two vortex-rings emerge from the two poles which are located at equal distances from centre. The value of $\lambda$ of this transition for $\eta=0.5$, is $\lambda_{t2(n=4)}=14.74$. As the diameter of these new vortex-rings increase with increasing $\lambda$, the two poles move toward the centre and join to each other at the centre at the critical point of  $\lambda=\lambda_{t3(n=4)}$. This transition which is a \textit{reverse type 1} transition, occurs at $\lambda_{t3(n=4)}=20.83$ for $\eta=0.5$. During this transition, the sign of the magnetic charge of the pole at the centre changes from negative to positive. For the interval of $\lambda_{t3(n=4)}<\lambda\leq 144$ the HEB solution includes a pole at the centre and two symmetric vortex-rings (figure \ref{fig.5}d).

Figures \ref{fig.5}a to \ref{fig.5}d, give a schematic illustration of the steps of the transitions along the HEB solution. This process is exactly the reverse of what happens along the fundamental solution of the case of $n=3$.

\begin{table}[tbh]
\small
\hskip-0.5in
\begin{tabular}{|c|ccccccccccc|}
\hline
  \multicolumn{12}{|c|} {Critical Points for the Case of~$n=4$}  \\ 
\hline

$\eta$ & 0&	0.1&	0.2&	0.3&	0.4&	0.5&	0.6&	0.7&	0.8&	0.9& 1\\ 
\hline
$\lambda$ (\textit{Type 3} Trans.)  & 3.04&	3.02&	2.98&	2.90&	2.80&	2.67&	2.50&	2.20&	1.85&	1.40&	0.80\\ 
\hline
$\lambda$ (Bifurcation)  & 6.207&	6.200&	6.17&	6.12&	6.06&	5.979&	5.87&	5.75&	5.61&	5.46&	5.28\\
\hline
$\lambda$ (\textit{Reverse Type 2}) & 14.97&	14.94&	14.91&	14.87&	14.82&	14.74&	14.65&	14.51&	14.34&	14.12&	13.77\\ 
\hline
$\lambda$ (\textit{Reverse Type 1}) & 21.23&	21.21&	21.16&	21.08&	20.97&	20.83&	20.66&	20.45&	20.21&	19.95&	19.65\\
\hline
\end{tabular}
\caption{\label{table.4} Table of the critical values of $\lambda$ for which the transitions of \textit{reverse type 1} and \textit{reverse type 2}, the transition of \textit{type 3} and bifurcation happen, for $n=4$.}
\end{table}

\begin{figure}[tbh]
\hskip-0.4in
 \includegraphics[width=6.6in,height=2.1in]{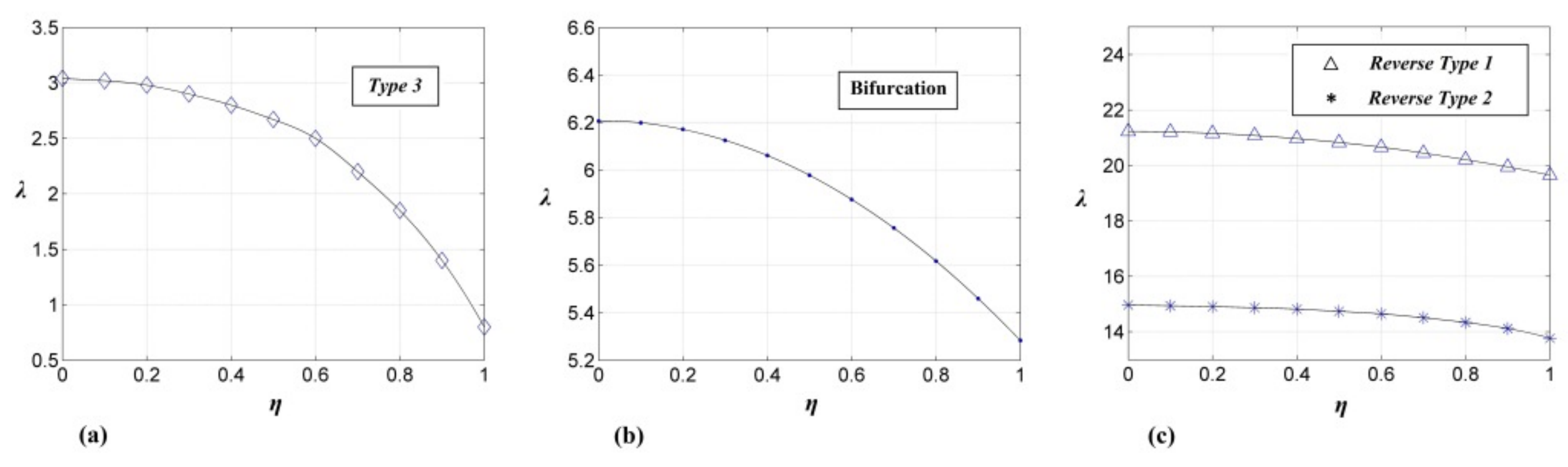} %\hskip-0.0in
	\caption{Higgs self-coupling constant $\lambda$, for transitions and bifurcation points versus the electric charge parameter $\eta$, where the cases of (a) the \textit{type 3} transition of the fundamental solution (b) the bifurcation point and (c) the two transitions of the HEB solution are shown when $n=4$.}
	\label{fig.6}
\end{figure}

The values of Higgs self-coupling constant, $\lambda$, for the critical points of the case of $n=4,$ for different values of $\eta$, are summarized in table \ref{table.4}. The sequence of the critical points for all values of $\eta$, is $\lambda_{t1(n=4)}<\lambda_{b(n=4)}<\lambda_{t2(n=4)}<\lambda_{t3(n=4)}$. The occurrence of these critical points versus the electric charge parameter, $\eta$, is illustrated in figure \ref{fig.6}.

\begin{table}[tbh]
\small
\hskip-0.55in
\begin{tabular}{|ccccccccccc|}
\hline
  \multicolumn{11}{|c|} {Fundamental Solution ~~($n=4, ~\eta=0.5$) }  \\ 
\hline
 & \multicolumn{4}{|c|} {1~Pole~and~3~Rings}&\multicolumn{6}{c|} {1~Pole~and~1~Ring}\\
\hline
$\lambda$ &\multicolumn{1}{|c} {0} &	0.01& 0.1	&	\multicolumn{1}{c|} {1}&	2.7&	10&	20&	50&	100&	144\\ 
\hline
$E$ &\multicolumn{1}{|c} {7.7240} &	9.2801&	11.2677&	\multicolumn{1}{c|} {14.9034}&	17.0431&	19.8432&	21.2034&	22.8692&	24.0344&	24.431\\ 
\hline
$D_{\rho1}$ & \multicolumn{1}{|c} {8.8473}&	5.8063&	4.7385&	\multicolumn{1}{c|} {4.0723}&	4.1217&	5.3083&	4.9858&	4.5245&	4.4591&	4.4922\\
\hline
$D_{\rho2}$ & \multicolumn{1}{|c} {10.2900}&	6.3740&	4.8220&\multicolumn{1}{c|} {4.1856}&-&-&-&-&-&-\\
\hline
$D_z$ & \multicolumn{1}{|c} {1.0820} &	0.5777&	0.3295&\multicolumn{1}{c|} {$<0.15$}&-&-&-&-&-&-\\
\hline
$Q$ & \multicolumn{1}{|c} {3.9209}&	2.4581&	1.7913&	\multicolumn{1}{c|} {1.2669}&	1.11&	1.0487&	1.0011&	0.9467&	0.9332&	0.9332\\ 
\hline
\multicolumn{11}{|c|} {3 Poles LEB Solution ~~($n=4, ~\eta=0.5$) }  \\ 
\hline

$\lambda$ &\multicolumn{1}{|c} {5.979} &	10&	16&	20&	26&	30&	40&	60&	100&	144\\ 
\hline
$E$ &\multicolumn{1}{|c} {19.4489} &	20.5771&	21.5615&	22.0195&	22.5483&	22.8313&	23.3853	&24.1295&	25.0081	&25.554\\ 
\hline
$d_z$ & \multicolumn{1}{|c} {1.5446}&	1.8339&	1.8018&	1.7767&	1.7473&	1.7322&	1.7041&	1.6695&	1.6398&	1.6223\\
\hline
$Q$ & \multicolumn{1}{|c} {1.0387}&	1.0317&	1.0051&	0.9932&	0.9804&	0.9741&	0.9628&	0.9500&	0.9380&	0.9309\\
\hline
\multicolumn{11}{|c|} {HEB Solution ~~($n=4, ~\eta=0.5$) }  \\ 
\hline
 & \multicolumn{2}{|c|} {3~Poles}&\multicolumn{2}{c|} {3~Poles~and~2~Rings}&\multicolumn{6}{c|} {1~Pole~and~2~Rings}\\
\hline
$\lambda$ &\multicolumn{1}{|c} {5.979} &\multicolumn{1}{c|} {10}	&	14.74&	\multicolumn{1}{c|} {20.83}&	26&	30&	40&	60&	100	&144\\ 
\hline
$E$ & \multicolumn{1}{|c} {19.4488}&	\multicolumn{1}{c|} {20.7114}&	21.7245&	\multicolumn{1}{c|} {22.6395}&	23.2248&	23.6001	&24.3449&	25.3633	&26.5737&	27.3734\\ 
\hline
$d_z$ & \multicolumn{1}{|c} {1.5424}&	\multicolumn{1}{c|} {0.7864}&	0.4749&	\multicolumn{1}{c|} {0.0154}&-&-&-&-&-&-\\
\hline
$D_z$ &\multicolumn{1}{|c} {-} &\multicolumn{1}{c|} {-}&0.5344&	\multicolumn{1}{c|} {0.4908}&	0.4666&	0.4618&	0.4328&	0.4308&	0.3699&	0.3458\\
\hline
$D_\rho$ &\multicolumn{1}{|c} {-}&\multicolumn{1}{c|} {-}& 0.5754&\multicolumn{1}{c|} {0.856}	&	0.9092&	0.9382&	0.9382&	0.9188&	0.8656&	0.8028\\
\hline
$Q$ & \multicolumn{1}{|c} {1.0385}&	\multicolumn{1}{c|} {0.9577}&	0.9233&	\multicolumn{1}{c|} {0.8992}&	0.8862&	0.8786&	0.8653&	0.8500&	0.8357&	0.8285\\
\hline
\end{tabular}
\caption{\label{table.5} Table of the dimensionless total energy $E$, the poles' separation $d_{z}$, the diameter of vortex-rings $D_{\rho}$, the distance of vortex-rings from $x$-$y$ plane $D_z$, and the electric charge $Q$, of different solutions, when $n=4$, $\eta=0.5$.($D_z<0.15$ for $1<\lambda<2.7)$}
\end{table}

\begin{figure}[tbh]
 \includegraphics[width=6in,height=4.1in]{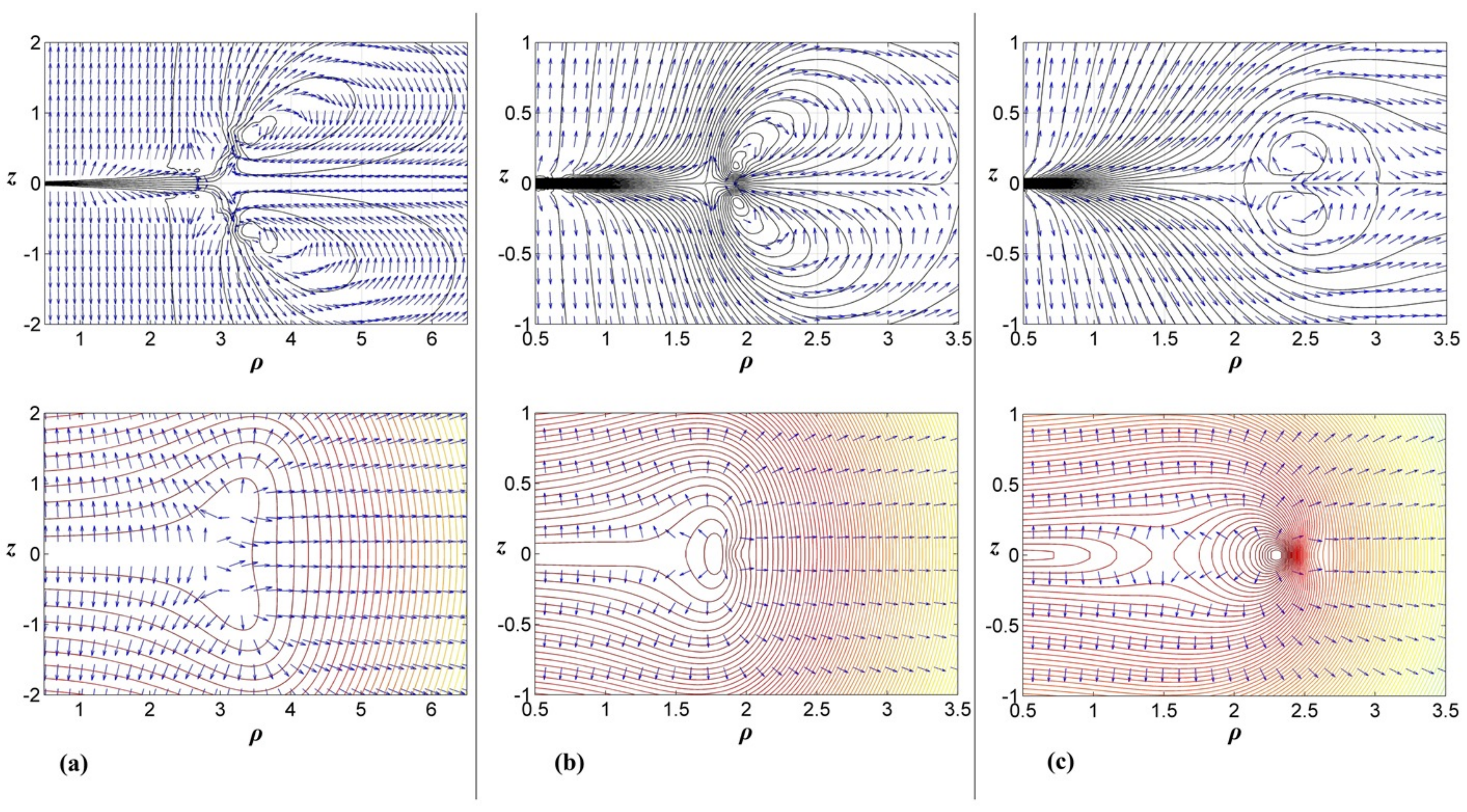} %\hskip-0.0in
	\caption{Magnetic field lines and magnetic field's unit vectors (top) and equipotential lines and unit vectors of electric field (bottom) of the fundamental solution for the case of $n=4$, $\eta=0.5$ where the cases of (a) $\lambda=0.01$, with three rings and one pole, (b) $\lambda=2$, with three rings and a pole and (c) $\lambda=4$, with a ring and a pole, after going through a \textit{type 3} transition, are shown.}
	\label{fig.7}
\end{figure}

\begin{figure}[tbh]
 \includegraphics[width=6in,height=4.1in]{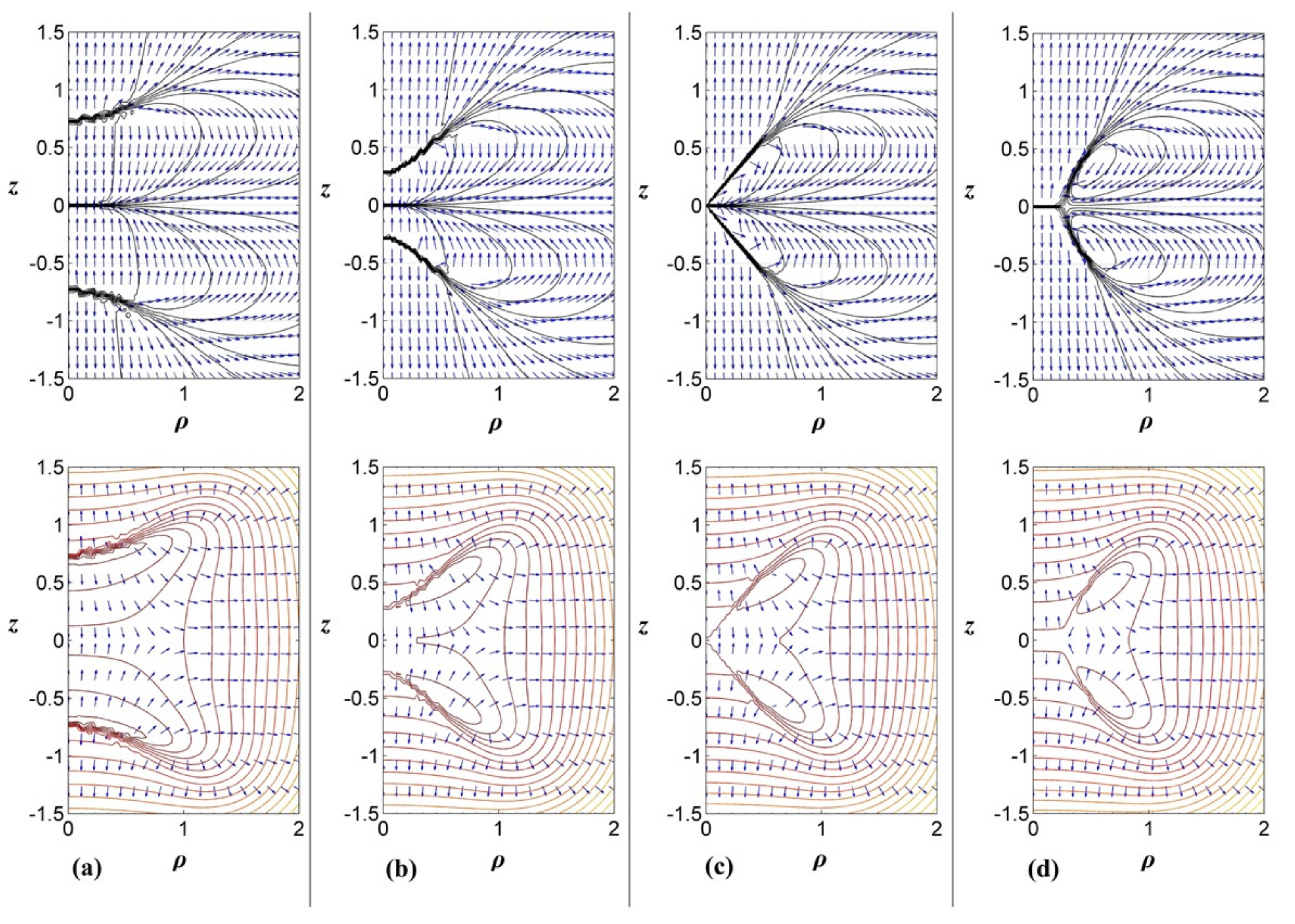} %\hskip-0.0in
	\caption{Magnetic field lines and magnetic field's unit vectors (top) and equipotential lines and unit vectors of electric field (bottom) of the HEB solution for the case of $n=4$, $\eta=0.5$. The cases of (a) $\lambda=10$, with three poles, (b) $\lambda=18$, with two rings and three poles (after of a \textit{reverse type 2} transition), (c) $\lambda=\lambda_{t2(n=4)}=20.83$, where the transition of \textit{reverse type 1} occurs and (d) $\lambda=30$, with one pole and two rings, are shown.}
	\label{fig.8}
\end{figure}

Figures \ref{fig.7} and \ref{fig.8} show the magnetic and electric field structures for the fundamental solution and the HEB solution respectively. The steps of the transitions are shown in these figures. As is illustrated in figure \ref{fig.7}, all the rings and also the pole at the centre, both before and after the transition, possess positive electric charges. 

Figure \ref{fig.8} however, indicates a change of electric charge for the pole which is located at the centre, during the transition of \textit{reverse type 1}. Integration on the small volume including the origin for HEB solution shows that the pole at the centre has a very small negative electric charge for the interval of $\lambda<\lambda_{t3(n=4)}$. The magnitude of this negative charge is very small in comparison with the magnitude of positive charge of each one of the other poles which are located on the symmetry axis. After the transition however, the pole which is located at the centre, acquires a small positive electric charge.

\begin{figure}[tbh]
	\includegraphics[width=6in,height=7in]{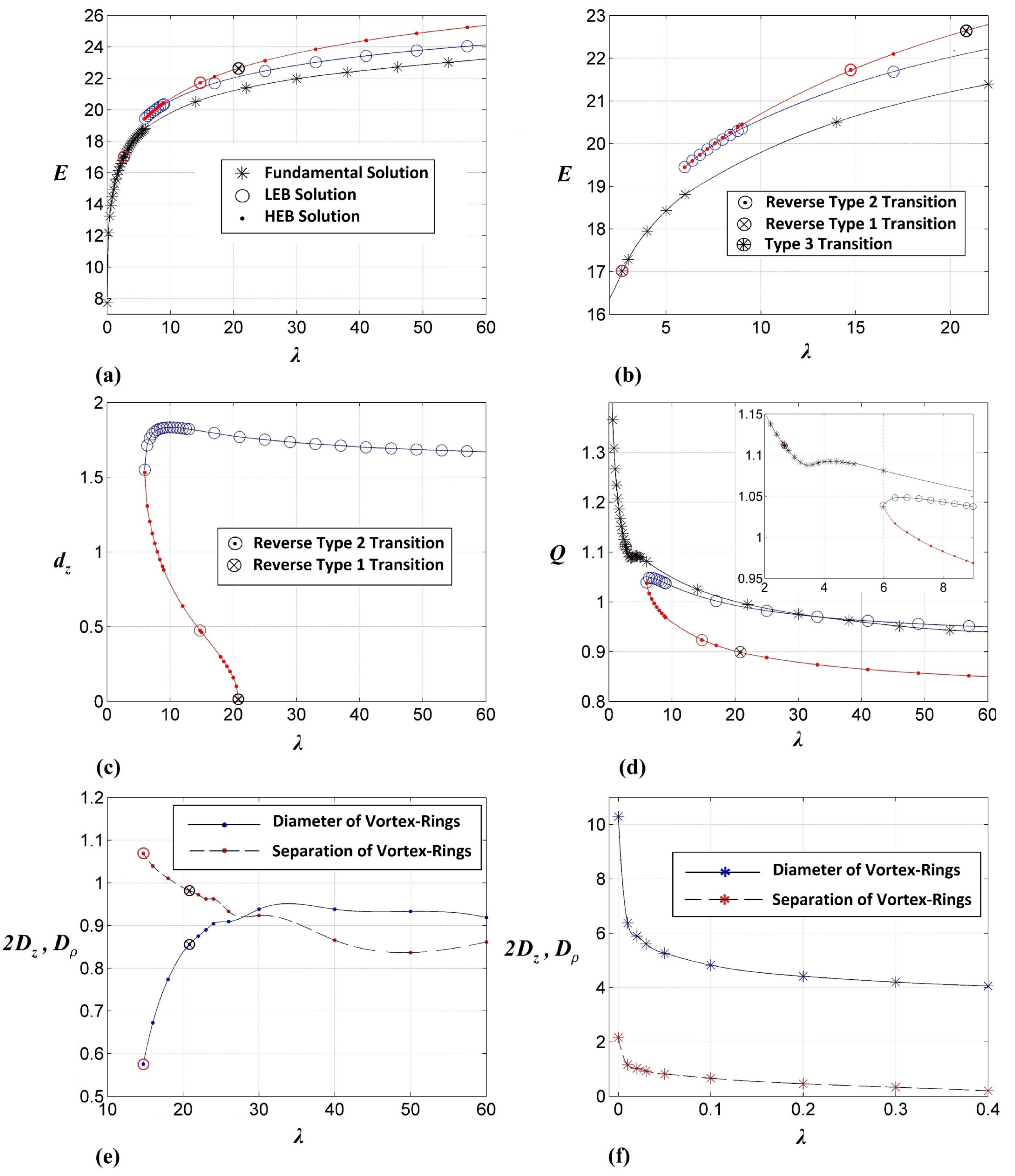} %\hskip-0.0in
	\caption{Plots of (a) and (b) the total energy, $E$, (c) the distance of the poles from the centre, $d_z$, (d) the total electric charge, $Q$,  and the separation of vortex-rings, $2D_z$, and diameter of vortex-rings, $D_\rho$,  for (e) the HEB case and (f)  the fundamental case, versus the Higgs self-coupling, $\lambda$, when $n=4$, $\eta=0.5$. }
	\label{fig.9}
\end{figure}

For the case of $n=4$ and $\eta=0.5$, the total electric charge of the fundamental solution has a local minimum value at $\lambda=3.468$ and a local maximum value at $\lambda=4.223$. The electric charge of the LEB solution also experiences a maximum value at $\lambda=6.561$. There is a crossover point of the electric charge of the fundamental and the LEB solutions at $\lambda=33.416$. Finally the separation of the poles of LEB solution reaches a maximum value in $\lambda=10.016$.
The details of the total energy, the total electric charge and geometrical properties of the solution are shown in figure \ref{fig.9}.\footnote{Our grid is not fine enough in some cases. For example, in the process of measuring the separation of the vortex-rings of thefundamental solution, the overlap of the vortex-rings makes it impossible to measure the separation in small distances. That is why the related diagram in figure \ref{fig.9}f, is plotted for a smaller interval.} More detailed quantitative information about this case is summarized in table \ref{table.5}.

\subsubsection{The $n=5$ case}
\label{sect:3.3.4}

This case also includes four critical points. The configuration of the fundamental solution consists of a pole at the centre and a vortex-ring on the $x$-$y$ plane for the interval of $0\leq \lambda\leq 300$. For small values of Higgs self-coupling constant and electrically neutral case, this solution has been obtained by Kleihaus et al. \cite{kn:7}. 

Bifurcations occur at higher energies. The first bifurcation takes place when $\lambda=\lambda_{b1(n=5)}$. We will refer to the lower and higher energy branches of this bifurcation as LEB1 and HEB1 respectively. This bifurcation for $\eta=0.5$ occurs at $\lambda_{b1(n=5)}=8.7$. The LEB1 possesses a three-poles configuration for the interval of $\lambda_{b1(n=5)}\leq\lambda\leq 300$. However, the HEB1 solution undergoes a \textit{reverse type 2} transition at $\lambda=\lambda_{t1(n=5)}$ and a \textit{reverse type 1} transition at $\lambda=\lambda_{t2(n=5)}$. The effects of these transitions are exactly like what occurs for HEB solution of $n=4$ case. For $\eta=0.5$, these transitions occur at $\lambda_{t1(n=5)}=29.31$ and $\lambda_{t2(n=5)}=33.8$. The schematic plot of the transitions are shown in figures \ref{fig.5}a to \ref{fig.5}d.

\begin{figure}[tbh]
\hskip-0.4in
 \includegraphics[width=6.6in,height=2.1in]{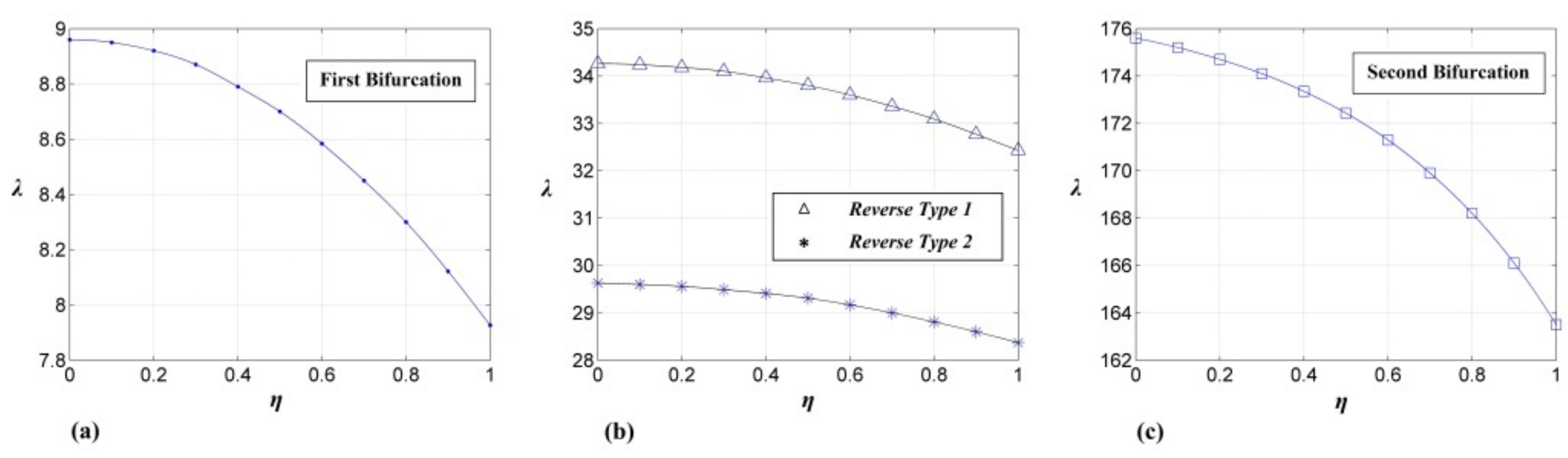} %\hskip-0.0in
	\caption{Higgs self-coupling constant $\lambda$, for the transition and the bifurcation points versus the electric charge parameter $\eta$, where the cases of (a) the first bifurcation point (b) the transitions of the HEB1 solution and (c) the second bifurcation point,  are shown when $n=5$.}
	\label{fig.10}
\end{figure}

\begin{figure}[tbh]
 \includegraphics[width=6in,height=4.1in]{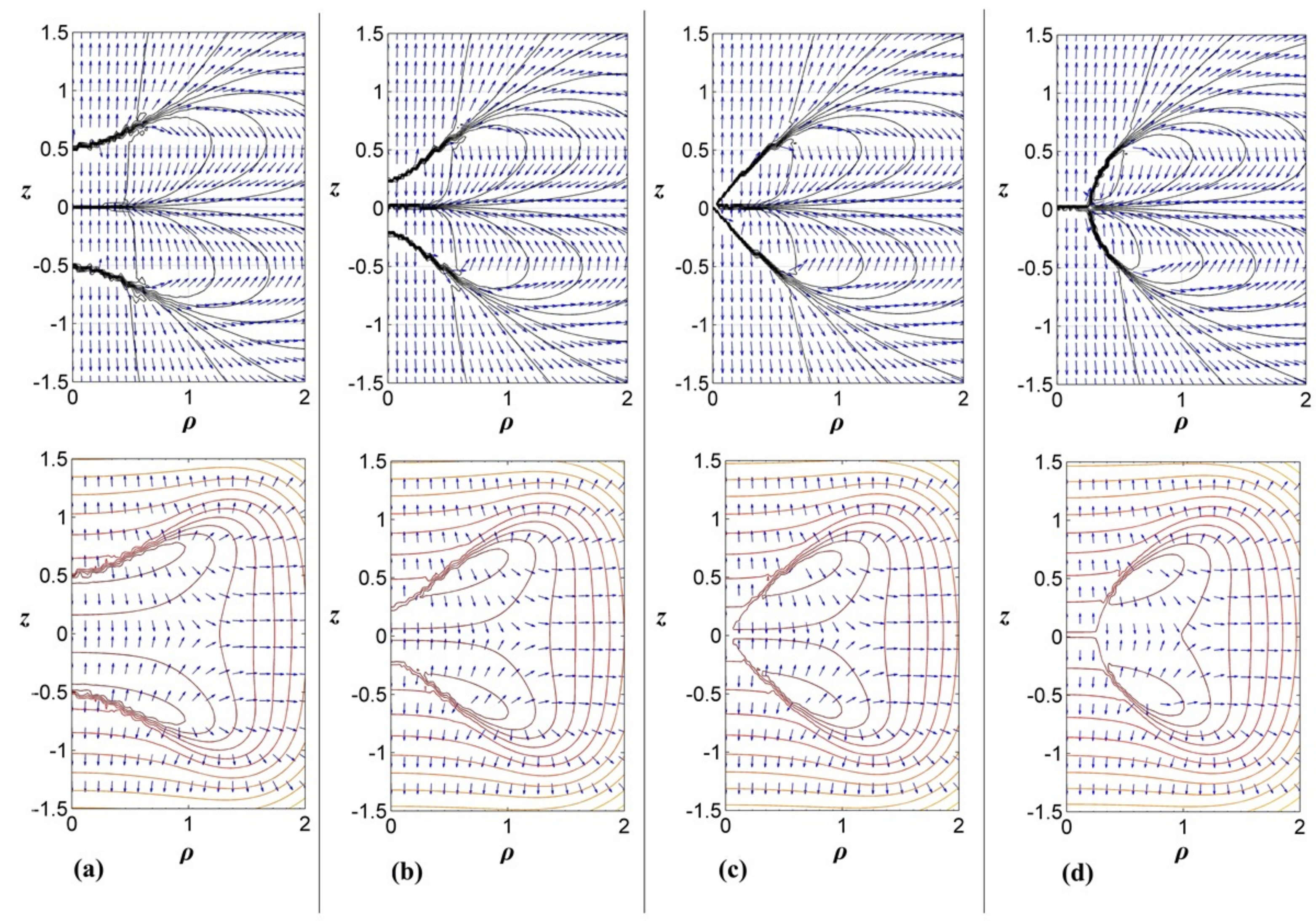} %\hskip-0.0in
	\caption{Magnetic field lines and magnetic field's unit vectors (top) and equipotential lines and unit vectors of electric field (bottom) of the HEB1 solution for the case of $n=5$ and $\eta=0.5$. The cases of (a) $\lambda=20$, where we have three poles, (b) $\lambda=30$, where there are three poles and two rings, (c) $\lambda=\lambda_{t2(n=5)}=33.8$, where the transition of \textit{reverse type 1} occurs and (d) $\lambda=50$, where we have a pole and two rings, are shown.}
	\label{fig.11}
\end{figure}

\begin{table}[tbh]
\small
\hskip-0.55in
\begin{tabular}{|c|ccccccccccc|}

\hline
  \multicolumn{12}{|c|} {Critical Points for the Case of~$n=5$}  \\ 
\hline

$\eta$ & 0&	0.1&	0.2&	0.3&	0.4&	0.5&	0.6&	0.7&	0.8&	0.9& 1\\ 
\hline
$\lambda$ (Bifurcation 1) & 8.960&	8.950&	8.921&	8.869&	8.790&	8.700&	8.584&	8.450&	8.300&	8.122&	7.927\\ 
\hline
$\lambda$ (\textit{Reverse Type 2})  & 29.63&	29.60&	29.56&	29.49&	29.41&	29.31&	29.17&	29.00&	28.81&	28.60&	28.37\\
\hline
$\lambda$ (\textit{Reverse Type 1}) & 34.26&	34.24&	34.18&	34.11&	33.96&	33.80&	33.60&	33.36&	33.09&	32.77&	32.43\\ 
\hline
$\lambda$ (Bifurcation~2) & 175.6&	175.2&	174.7&	174.1&	173.35&	172.43&	171.3&	169.9&	168.2&	166.1&	163.5\\
\hline

\end{tabular}
\caption{\label{table.6} Table of the critical values of $\lambda$ for which the transitions of \textit{reverse type 1} and \textit{reverse type 2} and the first and second bifurcation happen, for $n=5$.}
\end{table}

\begin{table}[tbh]
\small
\hskip-0.6in
\begin{tabular}{|c|cccccccccc|}
\hline
  \multicolumn{11}{|c|} {1~Pole~and~1~Ring~Fundamental Solution ~~($n=5, ~\eta=0.5$) }  \\ 
\hline

$\lambda$ & 0&	0.1&	1&	10&	50&	100&	150	&200&	250&300\\ 
\hline
$E$ & 9.0374&	13.7540&	18.3788	&24.2527&	27.9365&	29.0464&	29.5043&	29.7600&	29.9309	&30.0571\\ 
\hline
$D_\rho$ & 16.1253&	7.6482&	6.2730&	6.4493&	5.5321&	5.5545&	5.5636&	5.5669&	5.5685&	5.5695\\
\hline
$Q$ & 4.6016&	2.0146&	1.4264&	1.1894&	1.0726&	1.0697&	1.0695&	1.0693&	1.0690	&1.0691\\

\hline 
  \multicolumn{11}{|c|} {3~Poles~LEB1 Solution~~($n=5, ~\eta=0.5$)  }  \\ 
\hline

$\lambda$ & 8.7	&10&	20&	30&	50&	100&	150&	200&	250	&300 \\ 
\hline
$E$ & 26.0636&	26.4906&	28.5045&	29.6421&	31.0156	&32.7172&	33.6030&	34.1247	&34.4830&	34.7421 \\ 
\hline
$d_z$ & 1.4279&	1.599&	1.6213&	1.5807&	1.5333&	1.4875&	1.4859&	1.5192&	1.5319&	1.5376\\ 
\hline
$Q$ & 1.0907&	1.0957&	1.0579&	1.036&	1.0137&	0.9929&	0.9863&	0.9841&	0.9826&	0.9810 \\ 

\hline 
  \multicolumn{11}{|c|} {HEB1 Solution~~($n=5, ~\eta=0.5$) }  \\ 
\hline
& \multicolumn{2}{c} {3~Pole}&\multicolumn{2}{|c} {3~Poles~and~2~Ring}&\multicolumn{6}{|c|} {1~Pole~and~2~Rings}\\
\hline
$\lambda$ &  8.7&	10&\multicolumn{1}{|c} {29.31}	&	33.8&\multicolumn{1}{|c} {50}	&	100&	150	&200&	250&	300\\ 
\hline
$E$ & 26.0625&	26.4992&\multicolumn{1}{|c} {30.1511}	&	30.6387&	\multicolumn{1}{|c} {31.9511}	&34.1469&	35.3327&	36.1201&	36.6966&	37.1442 \\ 
\hline
$d_z $ & 1.4026&	1.1145&	\multicolumn{1}{|c} {0.2827}	&0.04&\multicolumn{1}{|c} {-}&-&-&-&-&- \\ 
\hline
$D_z$ & -&-&\multicolumn{1}{|c} {0.4747}&	0.4604&\multicolumn{1}{|c} {0.3902}	&	0.3342&	0.3077	&0.2883&	0.2738&	0.2641\\ 
\hline
$D_\rho$ & -&-&\multicolumn{1}{|c} {0.7394}&	0.7978&	\multicolumn{1}{|c} {0.8372}&	0.818&	0.78&	0.7414&	0.7074&	0.6784 \\ 
\hline
$Q$ & 1.0887&	1.0582&	\multicolumn{1}{|c} {0.9614}&	0.9535&	\multicolumn{1}{|c} {0.9349}&	0.9104&	0.9001&	0.8946&	0.8907	&0.8881\\ 

\hline 
  \multicolumn{11}{|c|} {3~Poles~LEB2 Solution~~($n=5, ~\eta=0.5$) }  \\ 
\hline

$\lambda$ &  172.5&	175	&180&	185&	190&	200&	225&	250&	275&	300\\ 
\hline
$E$ & 33.8403&	33.8679&	33.9127&	33.9587&	34.0040&	34.0915&	34.2897	&34.4612&	34.6107&	34.7422 \\ 
\hline
$d_z $ & 1.5889&	1.588&	1.5785&	1.5734&	1.5699&	1.5651&	1.5582&	1.5539&	1.551	&1.5489 \\ 
\hline
$Q$ & 0.9955&	0.9954&	0.9946&	0.9941&	0.9937&	0.9931&	0.9917&	0.9911&	0.9906&	0.9902\\

\hline 
  \multicolumn{11}{|c|} {3~Poles~HEB2 Solution~~($n=5, ~\eta=0.5$) }  \\ 
\hline
$\lambda$ &  172.5&	175	&180&	185&	190&	200&	225&	250&	275&	300\\ 
\hline
$E$ & 33.8432&	33.8722&33.9326	&	33.9876&	34.0391&	34.1339&	34.3367&	34.5024&	34.6406&	34.7578 \\ 
\hline
$d_z $ & 1.5912&	1.5937&1.6041	&	1.6100&	1.6140	&1.6194	&1.6271&	1.6314&	1.6340&	1.6358 \\ 
\hline
$Q$ & 0.9954&	0.9955&0.9958&	0.9958&	0.9957&	0.9956&	0.9943&	0.9939&	0.9933&	0.9925\\
\hline
\end{tabular}
\caption{\label{table.7} Table of the dimensionless total energy $E$, the poles' separation $d_{z}$, the diameter of vortex-rings $D_{\rho}$, the distance of vortex-rings from $x$-$y$ plane $D_z$, and the electric charge $Q$, of different solutions, when $n=5$, $\eta=0.5$.}
\end{table}

The second bifurcation happens at $\lambda=\lambda_{b2(n=5)}$. This critical point possesses a higher energy in comparison with the fundamental solution but its energy is slightly less than the energy of the LEB1 solution. The value of $\lambda$ for this bifurcation in case of $\eta=0.5$ is $\lambda_{t1(n=5)}=172.43$. We will refer to the branches of this new bifurcation as LEB2 and HEB2. Both of these new solutions have the three-poles structure within the interval of $\lambda_{b2(n=5)}\leq\lambda\leq 300$ and no transition occurs here. 

This means that we have only one solution for the interval of $0\leq \lambda< \lambda_{b1(n=5)}$. For the interval of $\lambda_{b1(n=5)}\leq \lambda< \lambda_{b2(n=5)}$, we have three distinct solutions and finally the number of solutions for the interval of $\lambda_{b2(n=5)}\leq \lambda\leq 300$, increases to five distinct solutions. The energy of LEB2 and HEB2 solutions are quite near to the energy of the LEB1 solution. The major guide for us to recognize these solutions as different is their different total electric charges. Indeed, without geometric analysis of these configurations and the detailed study of their electric charges, it was possible to assume the new branches as numerical errors around the LEB1 solution.\footnote{This shows that, it would be always useful to study the dyon cases instead of electrically neutral cases. The study of electric charge can help us to remove some possible degeneracies.}

Table \ref{table.6} includes the detailed information about the critical values of $\lambda$ in which the two bifurcations and the two transitions occur for different values of $\eta$, when $n=5$. The position of these critical points also are shown in figure \ref{fig.10}. The sequence of the critical points for this case is $\lambda_{b1(n=5)}<\lambda_{t1(n=5)}<\lambda_{t2(n=5)}<\lambda_{b2(n=5)}$.

\begin{figure}[tbh]
	\includegraphics[width=6in,height=7in]{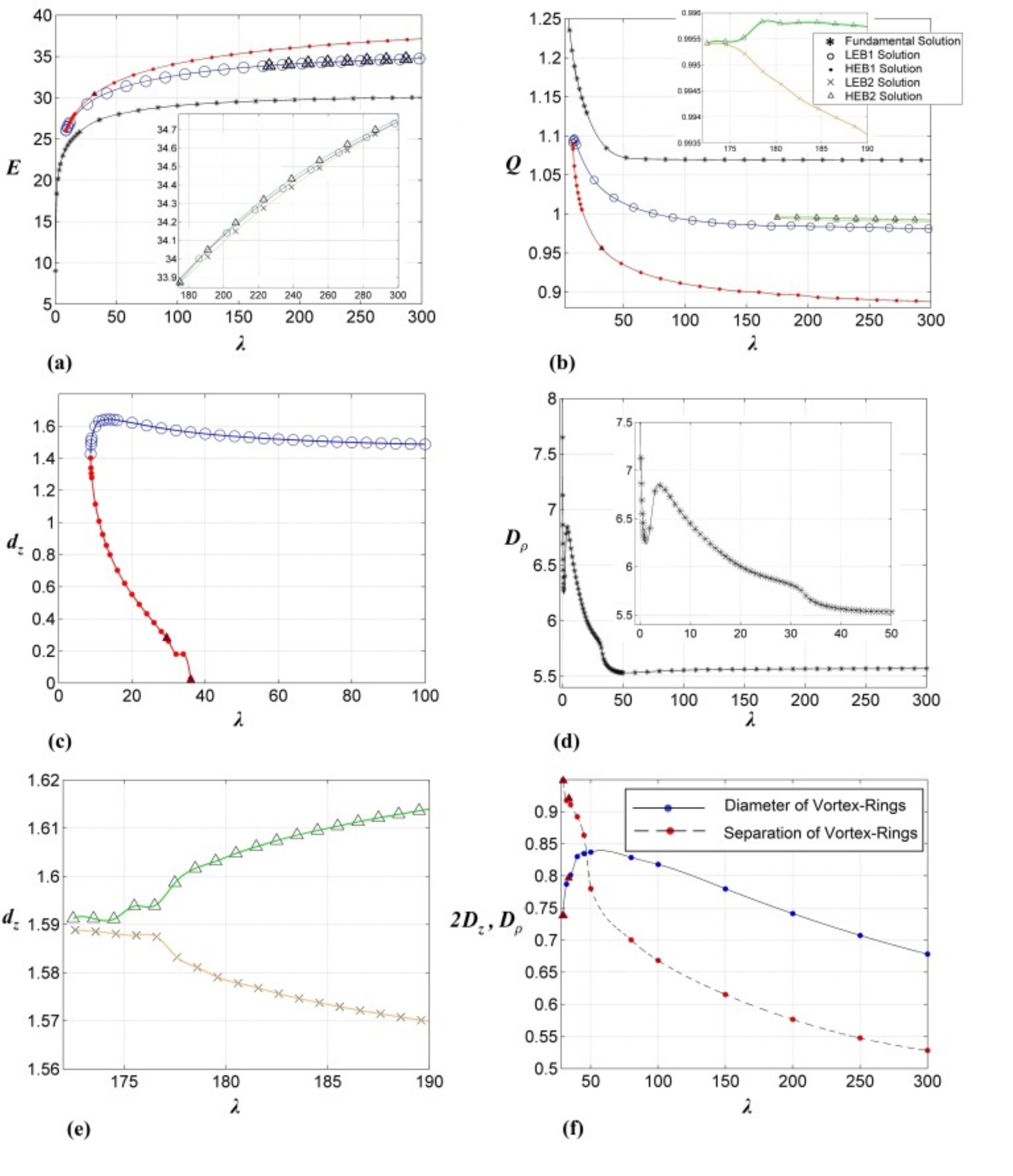} %\hskip-0.0in
	\caption{Plots of (a) the total energy, $E$, (b) the total electric charge, $Q$, (c) the distance of the poles from the centre, $d_z$, for LEB1 and HEB1, (d) diameter of vortex-ring, $D_\rho$ of the fundamental solution,  (e) the distance of the poles from the centre, $d_z$, for LEB2 and HEB2 and (f) the separation of vortex-rings, $2D_z$, and the diameter of vortex-rings, $D_\rho$, for the HEB1 solution, versus the Higgs self-coupling, $\lambda$, when $n=5$, $\eta=0.5$. The location of transitions (which are very close to each other) on HEB1, are shown with solid triangles. }
	\label{fig.12}
\end{figure}

The steps of the transitions along the HEB1 solution are shown in figure \ref{fig.11}. The direction of the magnetic field's unit vectors obviously shows that the sign of the magnetic charge at the centre changes from negative to positive during the transition of \textit{reverse type 1} at the critical point of $\lambda=\lambda_{t2(n=5)}$. Also, using the integration on the small volume including the origin, we can see that at the same critical point, the very small negative electric charge of the pole which is located at the centre, changes to a small positive charge. 

 For the fundamental solution, the diameter of the vortex-ring has a local minimum value at $\lambda=1.284$ and a local maximum at $\lambda=3.676$. For LEB1 solution, the total electric charge of the solution becomes maximum at $\lambda=9.328$ and the separation of the poles has a maximum at $\lambda=13.603$ and finally for the HEB1 solution, the diameter of the vortex-rings has a maximum value at the point of $\lambda=57.546$. The general form of the total energy, the total electric charge and geometrical properties of the solutions with respect to Higgs self-coupling is shown in figure \ref{fig.12}.\footnote{Near the second bifurcation point, the quality of convergence decreases rapidly. This is common for all kinds of the bifurcation points but it's more devastative for this bifurcation. So, the related values for geometrical properties of the system at these areas are not accurate and they have just indicative use.}  Also, table \ref{table.7} includes detailed data about each one of these five distinct solutions.

\section{Summary and Comments}
\label{sect:4}

The current study investigated the three-poles MAC system of solutions in the SU(2) YMH theory with net magnetic charge, $n$. For the first time, the presence of more than one bifurcation is shown in this paper. Also the presence of transitions in more than one of the solutions (branches) is introduced for the first time.

\begin{figure}[tbh]
\hskip-0.3in
\includegraphics[width=6.6in,height=2.2in]{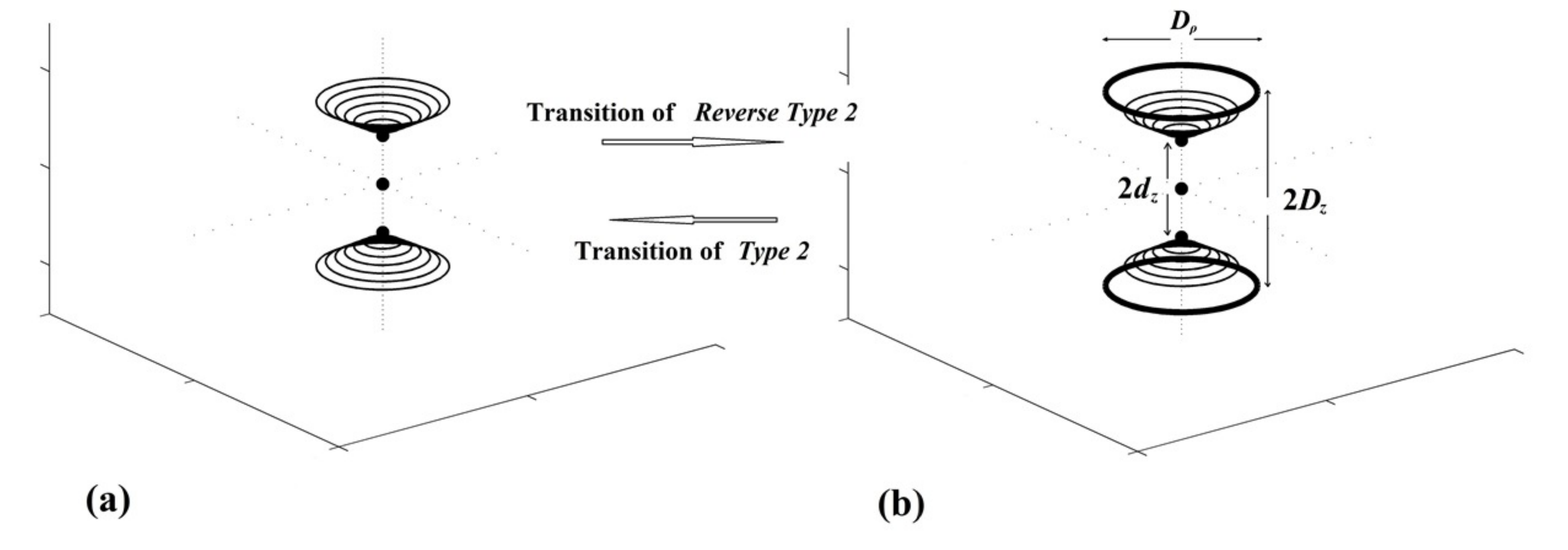} %\hskip-0.0in
\caption{A more detailed scheme for transition of \textit{type 2} (\textit{reverse type 2}). The tiny rings around the outer poles in three-poles configuration are shown.}
	\label{fig.13}
\end{figure}

\begin{figure}[tbh]
\hskip-0.3in
\includegraphics[width=6.6in,height=2.2in]{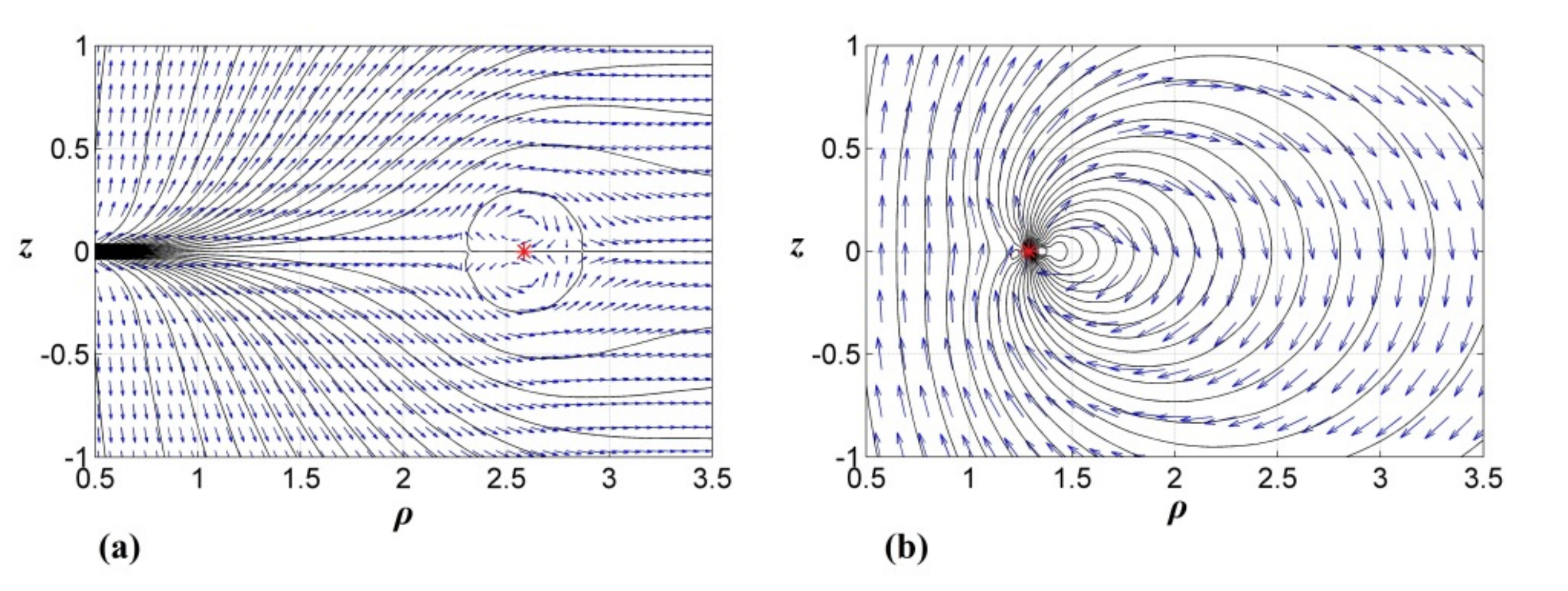} %\hskip-0.0in
\caption{Magnetic field lines and magnetic field's unit vectors for the case (a) a vortex-ring of the three-poles MAC system (Fundamental solution with $n=4, \eta= 0.5, \lambda=5$ and $D_\rho=5.178$) and (b) a vortex-ring of the MAP system (Fundamental solution with $n=3, \eta= 0.25, \lambda=30$ and $D_\rho=2.58)$. The asterisk shows the exact location of the vortex-ring. }
	\label{fig.14}
\end{figure}

\begin{figure}[tbh]
\hskip-0.3in
\includegraphics[width=6.6in,height=6in]{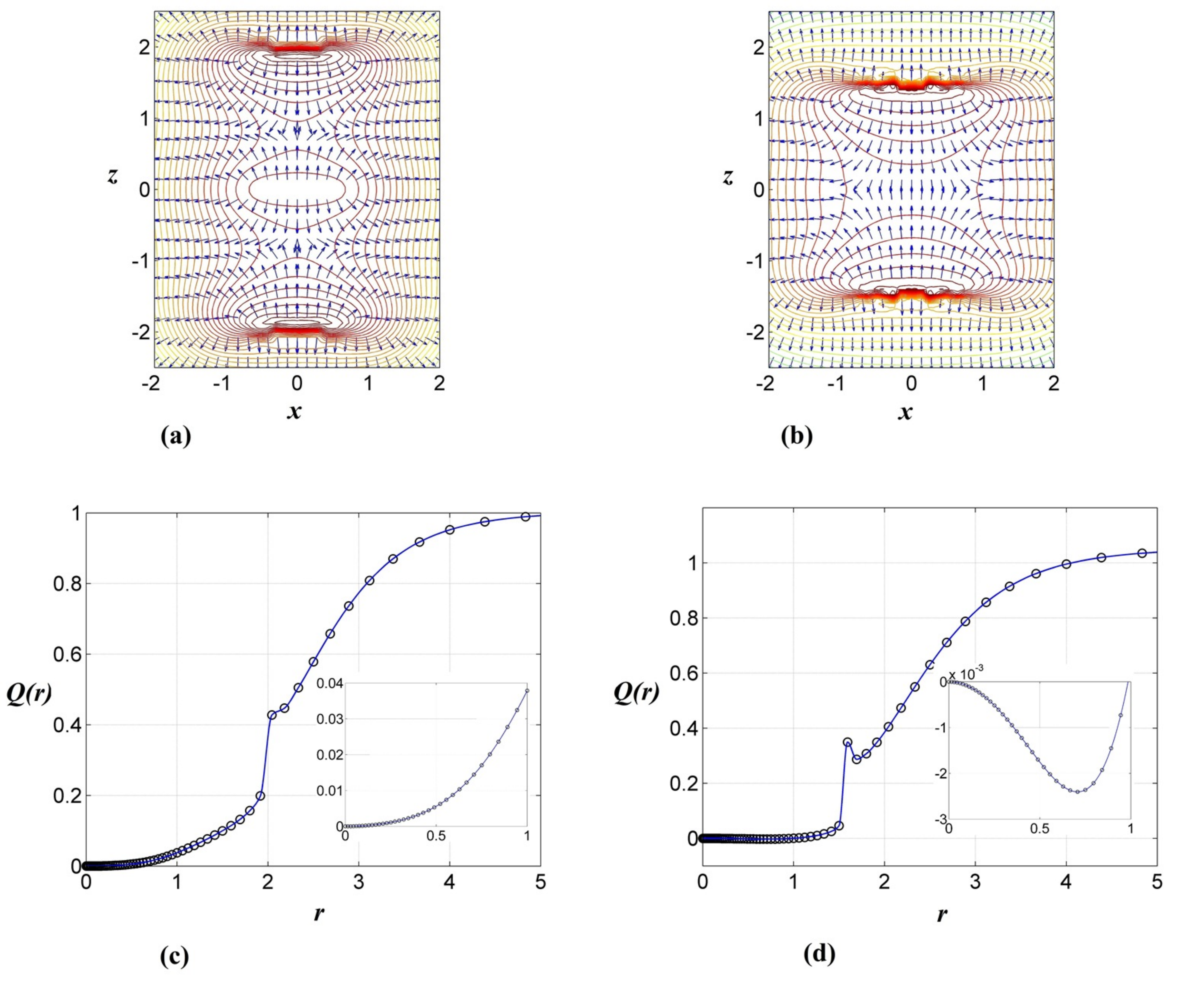} %\hskip-0.0in
\caption{Equipotential lines and unit vectors of electric field for three-poles configurations of (a) the HEB solution with $n=3, \eta= 0.5$ and $\lambda=5$ and (b) the LEB1 solution with $n=5, \eta= 0.5$ and $\lambda=30$. Integration on the volume including the origin shows that the total electric charge inside a sphere of radius $r$, $Q(r)$, at small radius has a small positive value for the case (a)(as is illustrated in (c)) and a very small negative value for the case (b) (as is illustrated in (d)).}
\label{fig.15}
\end{figure}

This study indicates that for the solutions with three isolated nodes on the symmetry axis (regardless of the presence or absence of vortex-rings) the outer poles are always encircled with tiny rings.\footnote{Kunz et al. have detected these rings for electrically neutral case with $n=3$ in ref. \cite{kn:8}.} The presence of these tiny rings makes it difficult to realize the accurate $\lambda$ for which the transition of \textit{type 2}(\textit{reverse type 2}) occurs. Figure \ref{fig.13} gives a more detailed schematic illustration of such a transition in presence of these tiny rings. Thus, to be truly accurate, in order to declare the occurence of a transition of \textit{reverse type 2}, in an inevitable way, we have to wait for a large enough value of $\lambda$ for which we can distinguish the new vortex-ring from the tiny rings around the outer poles. That is why as is shown in figure \ref{fig.13} b, for such a transition, we recorded $D_\rho>0$ and $2D_z>2d_z$.

For MAC system of the solutions with odd number of nodes on the symmetry axis, because of the symmetry of the magnetic charge with respect to the origin, the total magnetic dipole moment and therefore the intrinsic angular momentum of the system vanish. This form of the charge distribution is quite different than what we see in the MAP system of the solutions or those MAC systems which have an even number of nodes on the symmetry axis. For the case of even number of poles, the charge distribution of all solutions are such that the magnetic charge of the upper and the lower hemispheres are equal in magnitude but different in sign. For such a system, no pole is observed at the centre, but even for the vortex-rings which appear on $x$-$y$ plane, there is a positive magnetic charge distribution for upper hemisphere and a negative charge distribution for lower hemisphere.

\begin{figure}[tbh]
\includegraphics[width=6in,height=5in]{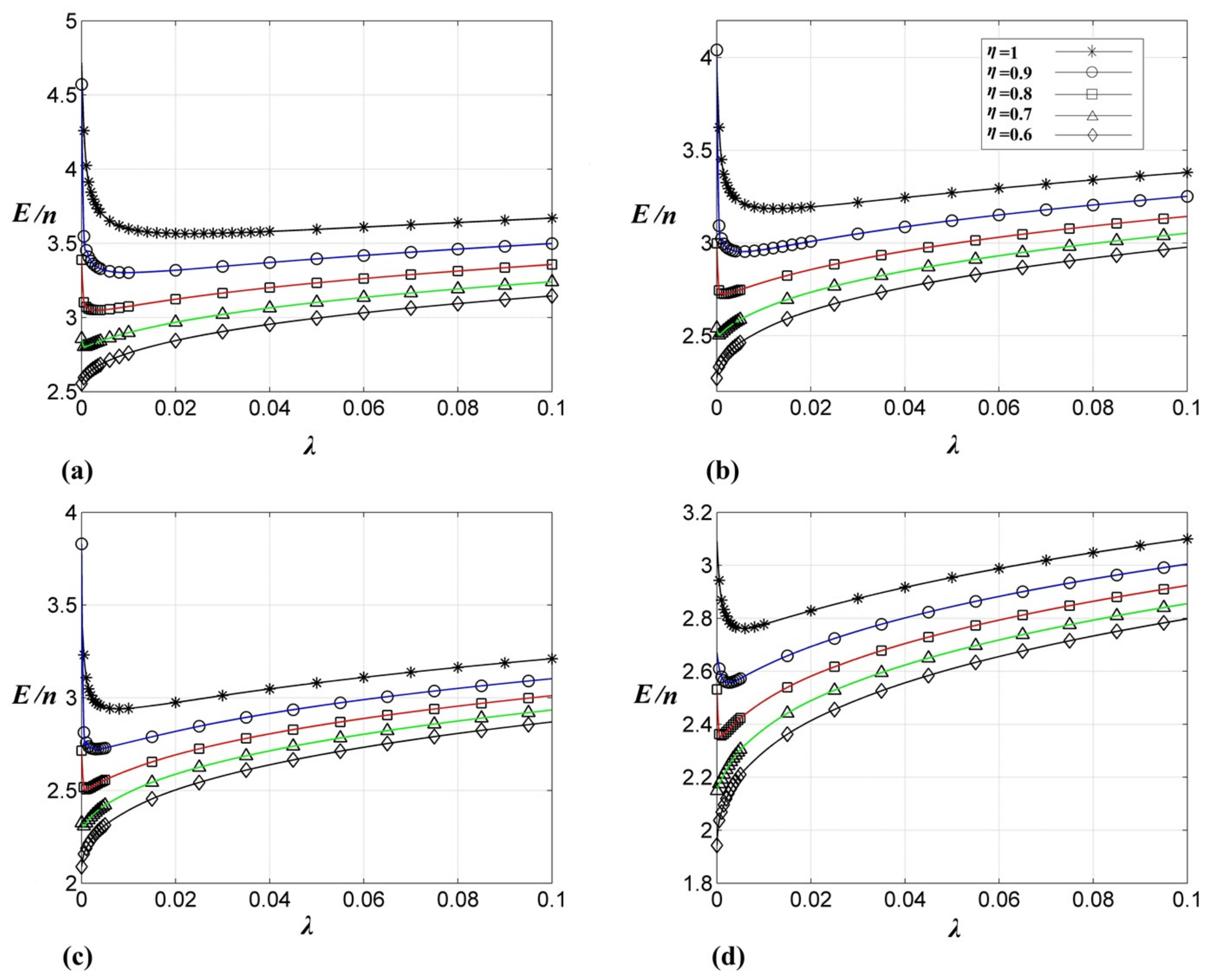} %\hskip-0.0in
\caption{Plots of the total energy per $n$, $E/n$, of the fundamental solutions versus the Higgs self-coupling, $\lambda$, for the cases of (a) $n=2$, (b) $n=3$, (c) $n=4$ and (d) $n=5$, and different values of $\eta$. }
	\label{fig.16}
\end{figure}
However as we mentioned above, in MAC cases with odd number of nodes on the symmetry axis, the charge distribution is symmetric with respect to the origin. This causes to have a new kind of magnetic charge distribution for the vortex-rings which are located on $x$-$y$ plane \cite{kn:7}. Figure \ref{fig.14} shows the difference in the orientation of the magnetic field's unit vectors between two different kinds of vortex-rings. As can be seen clearly, in the MAP case the upper hemisphere possesses positive electric charge and the lower hemisphere has negative charge. This is while for the vortex-ring of the three-poles MAC system, the negative charge is further from the centre in comparison with the positive charge.

Also the presence of a pole at the centre in MAC systems with the odd number of poles, causes another difference with those cases with even number of poles. For the systems with odd number of nodes, the sign of the magnetic charge of the pole which is located at the centre changes during the \textit{type 1} (or the \textit{reverse type 1}) transition whereas no such changes occur for the poles in the case of systems with even number of nodes. A similar phenomenon is detected for the electric charge of the pole which is located at the centre. The \textit{type 1} transition causes the sign of electric charge of the pole at the centre to change from positive to negative within a short interval of $\lambda$ (for $n=3$), and the \textit{reverse type 1} transition causes the negatively charged pole at the centre to acquire a positive electric charge.

This study shows that, considering the electric charge configuration, there are two major kinds of three-poles solutions. Integration over small volume including the origin shows that for the first kind, the pole which is located at the centre has a small positive electric charge (figures \ref{fig.15}a and \ref{fig.15}c ), while for the second kind, the pole which is located at the centre, has a very small negative electric charge (figures \ref{fig.15}b and \ref{fig.15}d ). The three-poles configurations of the LEB and the HEB solutions in case of $n=3$ are from first kind while the three-poles configurations of the HEB solution in the case of $n=4$ and the LEB1, the HEB1, the LEB2 and the HEB2 solutions in the case of $n=5$, are from the second kind. Figure \ref{fig.15} compares these two different kinds of three-poles configuration. In figures \ref{fig.15}c and \ref{fig.15}d, $Q(r)$ is the total electric charge inside the sphere of radius $r$ centred at origin. For the LEB solution of the case of $n=4$, an unexpected transformation from one kind to the other is detected. During this transformation, the negatively charged pole located at the centre, acquires a positive electric charge. This transformation for $\eta=0.5$, occurs at $\lambda=7.61$.

It is found that regardless of the value of $\phi$-winding number of the solutions, the electric charge of the fundamental solutions decreases rapidly within a very small interval of $0\leq\lambda\leq0.01$. At the same interval the total energy of the solution increases with increasing $\lambda$ (for $\eta<0.7$). The energy of the bifurcating branches are always larger than the energy of the fundamental solutions.
Also this study shows that, the critical points of transition, bifurcation and joining points for larger values of electric charge parameter, $\eta$, appear at smaller values of Higgs self-coupling constant, $\lambda$.

Figure \ref{fig.16} indicates that, for very small values of $\lambda$, the diagram of energy versus $\lambda$ has two different behaviours for the two cases of $\eta<0.7$ and $\eta>0.7$. Total electric charge, Q, is infinite for $\lambda=0$ and $\eta=1$. This fact causes the total energy to become infinite for $\lambda=0$ and $\eta=1$. For $0.7<\eta<1$ the total energy is not infinite at $\lambda=0$. However for those values of $\eta$, the total energy increases very fast as $\lambda$ decreases within the very small interval of $0\leq\lambda<10^{-3} $.   
%%%%%%%%%%%%%%%%%%%%%%%%%%%%%%%%%%%%%%%%%%%%%%% \section{Acknowledgement} %%%%%%%%%%%%%%%%%%%%%%%%%%%%%%%%%%%%%%%%%%%%%%%%%%%%%%%%%%%%%%%%
%\newpage

\section*{Acknowledgement}
The authors would like to thank Universiti Sains Malaysia for the RU research grant (Grant No. 1001/PFIZIK/811180).
%\newpage

\end{document}